\UseRawInputEncoding 
\documentclass{elsart}
\usepackage{graphicx}
\usepackage{url}
\usepackage{graphicx}
\usepackage{subcaption}
\usepackage{amsmath}

\begin{document}
\raggedbottom

\begin{frontmatter}

\title{Performance Analysis of UNet and Variants for Medical Image Segmentation}

\author[1]{Walid Ehab}
\author[1]{Yongmin Li}
\address[1]{Brunel University London,
  United Kingdom}

\maketitle

\begin{abstract}

Medical imaging plays a crucial role in modern healthcare by providing non-invasive visualisation of internal structures and abnormalities, enabling early disease detection, accurate diagnosis, and treatment planning. This study aims to explore the application of deep learning models, particularly focusing on the UNet architecture and its variants, in medical image segmentation. We seek to evaluate the  performance of these models across various challenging medical image segmentation tasks, addressing issues such as image normalization, resizing, architecture choices, loss function design, and hyperparameter tuning. The findings reveal that the standard UNet, when extended with a deep network layer, is a proficient medical image segmentation model, while the Res-UNet and Attention Res-UNet architectures demonstrate smoother convergence and superior performance, particularly when handling fine image details. The study also addresses the challenge of high class imbalance through careful preprocessing and loss function definitions. 
We anticipate that the results of this study will provide useful insights for researchers seeking to apply these models to new medical imaging problems and offer guidance and best practices for their implementation.

\end{abstract}


\end{frontmatter}





\section{Introduction}
Medical image segmentation is a critical aspect of medical image analysis and computer-aided diagnosis, involving the partitioning of images into meaningful regions for the identification of structures such as organs, tumors, and vessels. Deep learning, with its ability to automatically extract complex features from vast medical image datasets, presents a promising solution to enhance segmentation accuracy. However, challenges persist due to the diversity of medical domains, necessitating tailored approaches and evaluation metrics. 

This research's primary goal is to comprehensively study state-of-the-art deep learning methods, focusing on UNet \cite{ronneberger2015unet} and its variants, Res-UNet \cite{he2015deep} and Attention Res-UNet \cite{MAJI_attention}, renowned for their effectiveness in complex medical image segmentation tasks. 

The main objectives of this work are to apply the UNet model and its variants to a number of representative medical image segmentation problems, adapt different image pre-processing and model training techniques, identify appropriate performance metrics, and evaluate the performance of these models. Hopefully, the findings of this study will offer useful guidance to researchers when applying these models to new medical imaging problems.

The remainder of this paper is orgainised as follows. The problems of medical imaging and previous studies on segmentation, particularly medical image segmentation, are reviewed in Section 2. The details of UNet, its variants, and evaluation methods are discussed in Section 3. The applications of the above models to three problems of medical image segmentation, including brain tumor segmentation, polyp segmentation, and heart segmentation, are presented in Sections 4, 5, and 6, respectively. Finally, the findings and future work are presented in Section 7.

\section{Background}

Medical imaging has been widely employed by healthcare professionals for the evaluation of various anatomical structures. Medical image segmentation is the process of assigning labels to individual pixels within an image, thereby converting raw images into meaningful spatial data \cite{liu2020survey}. Currently, clinicians still largely perform this segmentation manually, a time-consuming process prone to both intra- and inter-observer variations \cite{haque2020deep}. The adoption of automatic segmentation methods holds significant promise, as it can enhance reproducibility and streamline clinical workflows. This is particularly relevant in the face of growing healthcare demands and a shortage of healthcare providers \cite{lock2022factors}.
The advance of new technologies has made it possible for automatic organ segmentation \cite{cootes1995active}, tumor segmentation \cite{litjens2017survey}, vessel segmentation \cite{fraz2012blood}, lesion detection and segmentation \cite{ren2015faster,valvano2019convolutional}, cardiac segmentation \cite{callaghan2015evaluation}, brain segmentation \cite{iglesias2011robust,havaei2017brain}, and bone segmentation \cite{fritscher2014automatic,ma2017cascade}, to name a few, in clinical practice.

Medical image segmentation is inherently influenced by the imaging modality employed. Computed Tomography (CT) imaging presents challenges related to similar tissue intensities, three-dimensional data, and radiation exposure control \cite{huda2003review}. Magnetic Resonance Imaging (MRI) introduces complexities in multi-contrast imaging, noise, and artifacts, as well as lengthy acquisition times \cite{zaitsev2015motion,plenge2012super}. Ultrasound imaging, although operator-dependent and prone to speckle noise, offers real-time imaging without ionizing radiation. Understanding the distinct characteristics and challenges of each modality is crucial for selecting appropriate segmentation techniques and optimizing the accuracy of medical image analysis \cite{rabbani2008speckle,wells2011medical,duarte2020speckle}.
Positron Emission Tomography (PET) imaging, commonly used for functional studies and cancer detection, faces resolution-noise trade-offs and requires advanced algorithms for accurate segmentation, distinguishing physiological from pathological regions \cite{kamiyoshihara2004congenital}.  X-ray imaging faces challenges due to the inherent two-dimensional projection of three-dimensional structures \cite{aichinger2012radiation}, making accurate segmentation difficult due to  overlapping structures and low contrast \cite{chen2018novel}.

Historically, image segmentation can be performed by using low-level image processing methods. For examples, thresholding is a straightforward technique that involves selecting a threshold value and classifying pixels as foreground or background based on their intensity values \cite{sezgin2004survey}.
Region-based segmentation methods focus on grouping pixels based on their spatial and intensity similarities \cite{felzenszwalb2004efficient}. The Watershed transform, introduced by Beucher and Serge\cite{beucher1992watershed}, is a region-based segmentation technique that has found applications in contour detection and image segmentation.

Statistical methods have also been developed for image segmentation. 
K-means clustering is a widely recognized method for partitioning an image into K clusters based on pixel intensity values \cite{jain1988algorithms}.
Active contours, often referred to as "snakes," were introduced by Kass, Witkin, and Terzopoulos\cite{kass1988snakes}.
Probabilistic modelling for medical image segmentation was presented in  \cite{Kaba:his2013,Kaba:hiss2014,Kaba:oe2015} where the Expectation-Maximisation process is adopted to model each segment as a mixture of Gaussians. 
The graph cut method utilises graph theory to partition an image into distinct regions based on pixel similarities and differences 
\cite{Dodo:bioimaging2018,Dodo:bioimaging2018_2,Salazar:jbhi2014,Salazar:his2012,Salazar:jaiscr2012,Salazar:icarcv2010,Salazar:cimi2011}.
The Markov Random Field (MRF) was adopted in 
\cite{Eltayef:ida2017,eltayef2016detection1,eltayef2016detection2,Eltayef:cbms2017}
for lesion segmentation in dermoscopy images in combination with  particle swarm optimisation, and  
for optic disc segmentation \cite{Salazar:his2012} and choroidal layer segmentation \cite{Wang:jbhi2017}.
The level-set method, based on partial differntial equations (PDE) , progressively  evaluates the differences among neighbouring pixels to find object boundaries and evolves contours to delineate regions of interest 
\cite{Dodo:jms2019,Dodo:best2019,Dodo:cbms2017,Dodo:bioimaging2019,Dodo:cbms2019,Dodo:access2019,Wang:jms2015,wang2020blood,Wang:icig2015,Wang:icig2015_2,Wang:jbhi2017}.

Over the past decade, the Deep Learning (DL) techniques stand as the cutting-edge approach for medical image segmentation. 
The Convolutional Neural Networks (CNN)
are inherently suited for volumetric medical image segmentation tasks. They can be customized by adjusting network depth and width to balance between computational efficiency and segmentation accuracy. Ensembling multiple 3D CNNs with diverse architectures has been effective in improving robustness and generalization to different medical imaging modalities \cite{drozdzal2018learning}.
Fully Convolutional Networks (FCN)
have been successfully adapted to medical image segmentation tasks by fine-tuning pre-trained models or designing architectures tailored to specific challenges. In scenarios where anatomical structures exhibit varying shapes and appearances, FCNs can be modified to include multi-scale and skip connections to capture both local and global information\cite{litjens2017survey}. 

The UNet  \cite{ronneberger2015unet} represents the most widely embraced variation among DL networks, featuring a U-shaped architecture with skip connections that enables the accurate delineation of objects in images \cite{litjens2017survey}.
SegNet, an encoder-decoder architecture, offers adaptability to various medical imaging modalities. Its encoder can be customized to incorporate domain-specific features, such as texture and intensity variations present in medical image\cite{badrinarayanan2017segnet}. Additionally, the decoder can be modified to handle the specific shape and structure of objects within the medical images, ensuring precise segmentation\cite{kao2019brain}. 

ResUNet \cite{he2015deep} extends the UNet architecture by introducing residual connections, which enable the network to train effectively, even with a large number of layers, thereby improving its ability to capture complex features in medical images. The integration of residual blocks in ResUNet facilitates the training of deeper networks and enhances segmentation accuracy, making it a valuable choice for tasks demanding the precise delineation of anatomical structures in medical image analysis.
Attention ResUNet \cite{MAJI_attention} builds on the ResUNet framework by incorporating attention mechanisms, allowing the network to selectively focus on informative regions in the input image while suppressing noise and irrelevant features. By introducing self-attention or spatial attention modules, Attention ResUNet enhances its segmentation capabilities, particularly in scenarios in which fine details and subtle variations in medical images are critical for accurate segmentation and diagnosis.

Recently, the nnUNet automatic segmentation framework, with its self-configuration mechanism taking into consideration of both cmoputer-hardware capabilities and dataset specific properties, has demonstrated segmentation performance that matches or closely approaches the state-of-the-art, as indicated in a study \cite{isensee2021nnu}. Extended models of nnUNet have been reported in \cite{mcconnell2022integrating,mcconnel2023,ndipenoch2022simultaneous,ndipenoch2023retinal} for various medical imaging applications.

The exploration of traditional image segmentation methods has revealed their strengths and limitations in simpler tasks but exposed vulnerabilities in complex medical imaging. General segmentation techniques adapted for medical applications, such as the Watershed transform and active contours, have shown promise in specific areas but come with their own limitations. The various domains of medical image segmentation, each with its unique challenges, highlight the complexity of this field. These challenges range from organ shape variability to tumor heterogeneity and vessel intricacies. In light of these challenges, the importance of UNet and its variants becomes evident. These deep learning approaches offer the potential to overcome the limitations of traditional methods, promising more accurate and adaptable segmentation solutions for complex medical images. Exploring UNet and its variants signifies a journey into harnessing the power of deep learning to address the intricacies of medical image segmentation. This endeavor seeks not only to understand the foundations of UNet but also to explore its potential in overcoming the limitations of traditional methods. Ultimately, this exploration aims to advance medical image analysis, leading to improved healthcare quality and patient outcomes in this critical field.







\section{Methods}


An overview of the deep learning models, including UNet, Res-UNet, and Attention Res-UNet, is provided in this section with details of the network architectures, filters of individual layers, connections between layers, specific functional mechanisms such as attention, activation functions, and normalisation. 

\subsection{UNet}
\begin{figure}[h]
    \centering
    \begin{subfigure}[b]{\linewidth}
        \centering
        \includegraphics[width=0.8\textwidth]{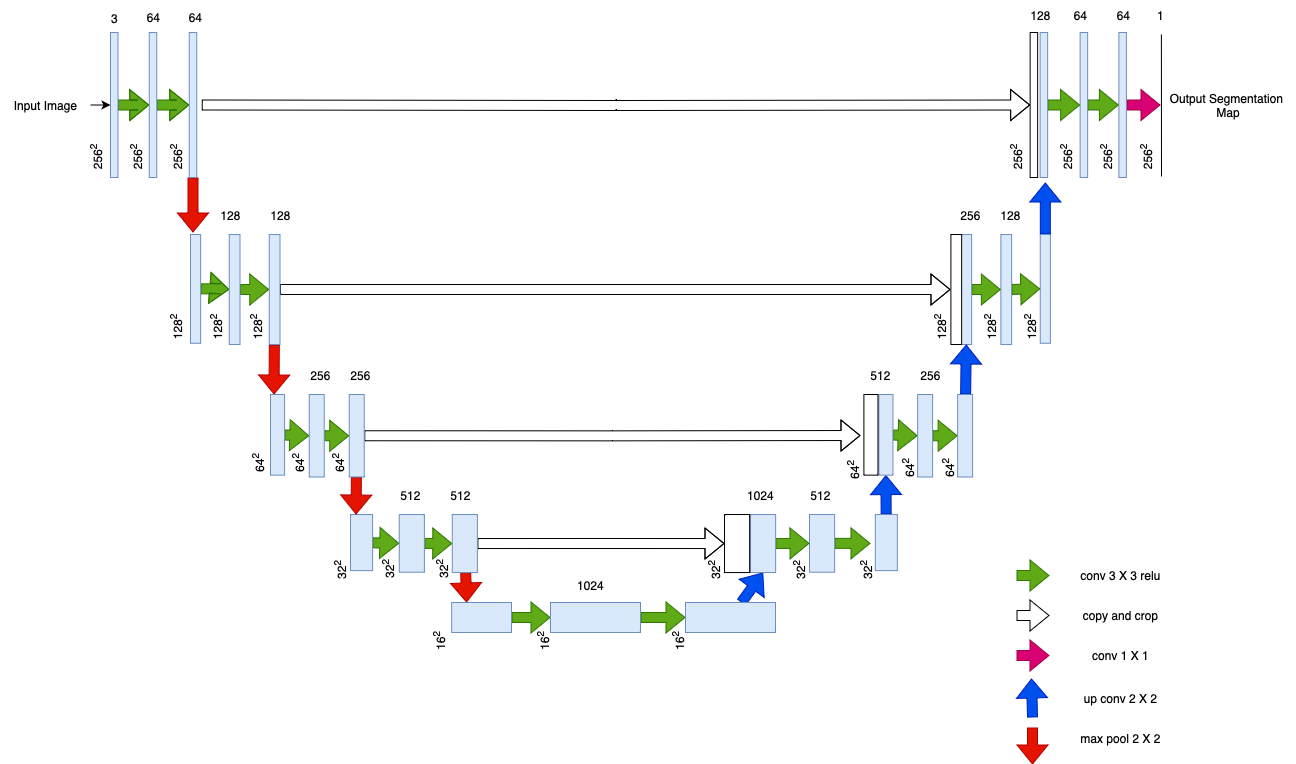}
        \caption{}
        \label{unet_proposed_brain}
    \end{subfigure}
    \begin{subfigure}[b]{\linewidth}
        \centering
        \includegraphics[width=0.8\textwidth]{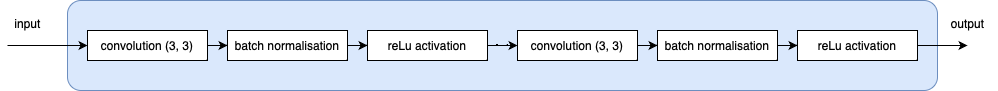}
        \caption{}
        \label{conv_block}
    \end{subfigure}
    \caption{UNet. (a) Network architecture. (b) Details of the convolution block.}    
\end{figure}

UNet is a convolutional neural network (CNN) architecture that was originally designed for biomedical image segmentation but has found applications in a wide range of image analysis tasks. Introduced by Ronneberger et al. in 2015 \cite{ronneberger2015unet}, UNet's architecture is characterized by its unique encoder-decoder structure and skip connections. Figure \ref{unet_proposed_brain} shows the general UNet architecture adopted for this project.

UNet's architecture consists of two main components: the contracting path (encoder) and the expansive path (decoder). This design enables UNet to capture both global and local features of the input image, making it highly effective for segmentation tasks.

\noindent
\textbf{Contracting path (Encoder):}
The contracting path is responsible for feature extraction. The UNet model built in this project has four encoding layers. Each encoding layer consists of 2 convolution layers or one convolution block, each followed by batch normalisation layers for ensuring normalisation, and a relu activation layer as shown in Fig \ref{conv_block}. The output from the convolution block is then passed through a a down sampling layer with max-pooling to reduce the spatial dimensions of the feature maps. The contracting path is crucial for building a rich feature representation.
After the four encoding layers the output passes through a bottleneck layer and then the upsampling layers(decoders).

\noindent
\textbf{Expansive path (Decoder):}
The expansive path aims to recover the original resolution of the image. The UNet model has four decoding layers. It comprises up-sampling and transposed convolutional layers. Importantly, skip connections connect the encoder and decoder at multiple levels. These skip connections allow the decoder to access feature maps from the contracting path, preserving spatial information and fine details.

\noindent
\textbf{Skip connections:}
Skip connections are a key innovation in UNet's architecture. They address the challenge of information loss during up-sampling. By providing shortcut connections between corresponding layers in the encoder and decoder, skip connections enable the model to combine low-level and high-level features effectively. This ensures that fine details are retained during the segmentation process.

\noindent
\textbf{Kernel size and number of filters:}
Throughout the structure, a kernel size of 3 is maintained for the convolution layers, as this filter size is common in image segmentation tasks. Smaller filter sizes capture local features, while larger filter sizes capture more global features. The number of filters in the first layer is set to 64. This is a common practice to start with a moderate number of filters and gradually increase the number of filters in deeper layers. It allows the network to learn hierarchical features. 

\noindent
\textbf{Final Fully Connected Convolutional layer:}
The output passes through a final fully connected convolution layer after four decoding layers. The size of kernel in the last layer depends on the number of classes(labels) present in the mask and is therefore tailored to needs of the tasks. The output from the convolutional layer passes through an activation function to produce the final output. The final activation function used also depends on the number of labels in the output. Final Kernel size and activation layer is mentioned for each task in the following chapters.

UNet's design makes it particularly effective for tasks where precise localization and detailed segmentation are required, such as medical image segmentation.

\subsection{Res-UNet}
\begin{figure}[h]
    \begin{subfigure}{\linewidth}
    \centering
    \includegraphics[width=0.8\textwidth]{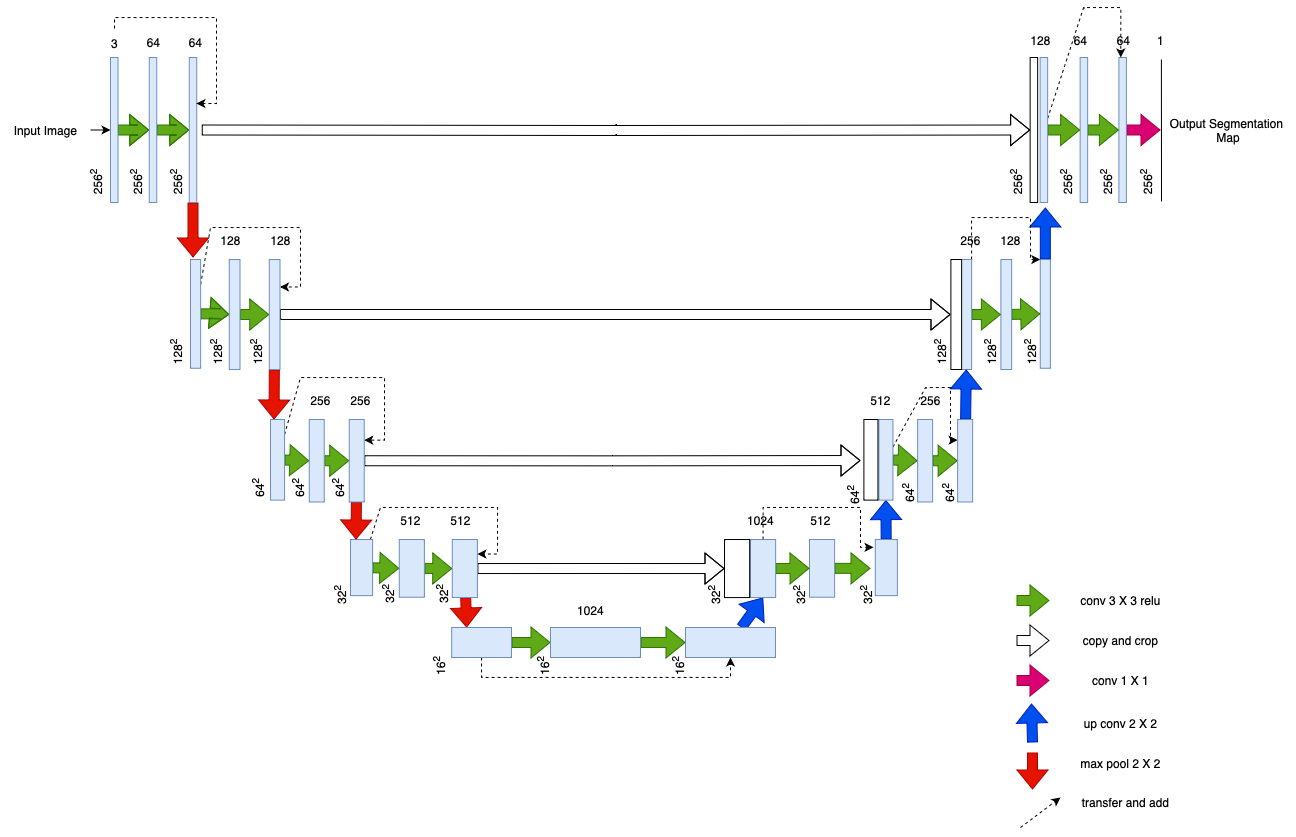}
    \caption{}
    \label{res_unet_proposed_brain}
    \end{subfigure}

    \begin{subfigure}{\linewidth}
    \centering
    \includegraphics[width=0.8\textwidth]{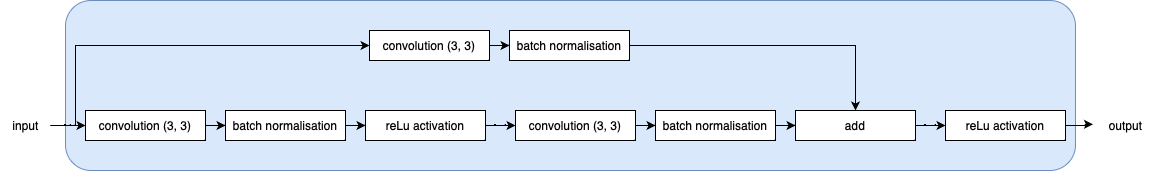}
    \caption{}
    \label{res_block}
    \end{subfigure}

    \caption{Res-UNet. (a) Network architecture; (b) Details of the residual convolution block.}

\end{figure}

Res-UNet is an extension of UNet that incorporates residual connections. Residual connections were introduced in the context of residual networks (ResNets)\cite{he2015deep} to address the vanishing gradient problem in deep networks. Res-UNet combines the strengths of UNet with the benefits of residual connections. The convolution block in UNet is replaced here with residual blocks which introduces an addition layer between the input at each block and the output from the last 3X3 convolutional block. 

\noindent
\textbf{Residual Connections:} Res-UNet incorporates residual connections between layers. These connections allow gradients to flow more easily during training, enabling the training of deeper networks without suffering from vanishing gradients.

\noindent
\textbf{Enhanced Information Flow:} The use of residual connections enhances the flow of information through the network, enabling it to capture long-range dependencies and complex structures in medical images.

The Res-UNet model adopted in this project has four encoding and four decoding layers. The overall architecture of the Res-UNet model and the residual convolutional block is provided in \ref{res_unet_proposed_brain} and \ref{res_block}. Res-UNet is known for its ability to handle deeper networks, which can be advantageous for capturing intricate details in medical images.

\subsection{Attention Res-UNet}
\begin{figure}[h]
    \centering
    \includegraphics[width=0.8\textwidth]{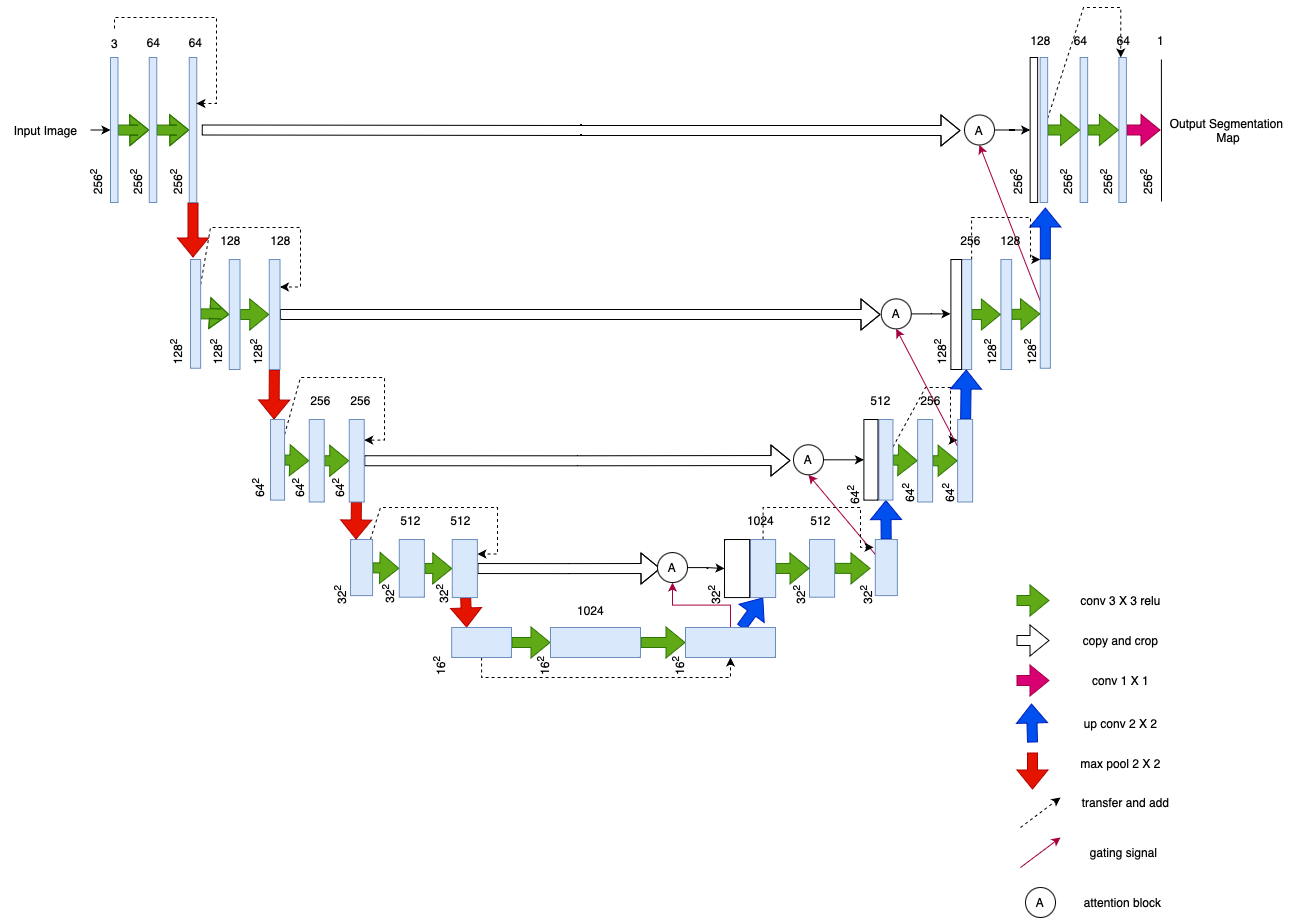}
    \caption{Proposed Attention Res-UNet Architecture}
    \label{att_res_unet_proposed_brain}
\end{figure}
Attention Res-UNet model\cite{MAJI_attention} builds on the Res-UNet architecture but introduces attention mechanisms. This is achieved through gating signals, which brings output from the lower layer to match the same dimension as the current layer, and an attention block, which combines information from two sources: the input feature map (x) and the gating signal (gating) to compute attention weights that determine how much focus or importance should be given to different spatial regions of the input feature map. Attention mechanisms enable the network to focus on salient regions of the input, improving its ability to differentiate between important and less important features. The key steps taken to implement the two blocks are explained below:\\

\noindent
\textbf{Gating Signal:}
The gating signal is a subnetwork or a set of operations employed to modulate the flow of information in an attention mechanism. In this specific implementation, the gating signal is generated as follows:
\begin{itemize}
    \item \textbf{Convolutional Layer:} A convolutional layer is used to transform the input features into a format compatible with the requirements of the attention mechanism. It adjusts the feature dimensionality if necessary.
    \item \textbf{Batch Normalization (Optional):} An optional batch normalization layer is applied to ensure that the output of the convolutional layer is well-scaled and centered, thereby aiding in stabilizing training.
    \item \textbf{ReLU Activation:} The ReLU activation function introduces non-linearity to the gating signal, helping capture complex patterns and relationships in the data.
\end{itemize}

    
    
    
    
    
    
\noindent
\textbf{Attention Block:} The attention block is a critical part of attention mechanisms employed in neural networks. Its primary purpose is to combine information from two sources—the input feature map (\(x\)) and the gating signal (\textit{gating}). Here's a breakdown of its functionality:
\begin{itemize}
    \item \textbf{Spatial Transformation (Theta\_x):} The input feature map (\(x\)) undergoes spatial transformation using convolutional operations. This transformation ensures that the feature map aligns with the dimensions of the gating signal.
    \item \textbf{Gating Signal Transformation (Phi\_g):} Similarly, the gating signal is subjected to transformation via convolutional operations to ensure appropriate spatial dimensions.
    \item \textbf{Combining Information:} The transformed gating signal (\textit{Phi\_g}) and the spatially transformed input feature map (\textit{Theta\_x}) are combined to capture relationships between different parts of the input.
    \item \textbf{Activation (ReLU):} The ReLU activation function is applied to the combined information, introducing non-linearity and enabling the capture of complex relationships.
    \item \textbf{Psi and Sigmoid Activation:} The combined information is further processed to produce attention weights (\textit{Psi}) using convolutional layers and a sigmoid activation. The sigmoid activation ensures that the attention weights are within the range of 0 to 1, indicating the degree of attention assigned to each spatial location.
    \item \textbf{Upsampling Psi:} The attention weights are upsampled to match the spatial dimensions of the original input feature map, ensuring alignment with the input.
    \item \textbf{Multiplication (Attention Operation):} The attention weights are multiplied element-wise with the original input feature map (\(x\)). This operation effectively directs attention to specific spatial locations in the feature map based on the computed attention weights.
    \item \textbf{Result and Batch Normalization:} The final result is obtained by applying additional convolutional layers and optional batch normalization, ensuring that the output is appropriately processed.
\end{itemize}
The gating signal prepares a modulating signal that influences the attention mechanism in the attention block. The attention block computes attention weights to focus on relevant spatial regions of the input feature map, which is particularly useful in tasks requiring fine-grained detail capture, such as image segmentation or object detection. The attention mechanism aids the network in prioritizing and weighting different spatial locations in the feature map, ultimately enhancing performance.

\subsection{Evaluation Methods}
The following metrics were adopted to evaluate the performance of the models:

\noindent
\textbf{Execution time: }Execution time is recorded for the training of each model. This is done to understand how long a model takes to converge. This is implemented using \verb|datetime| library in python.

\noindent
\textbf{Validation Loss over Epochs}: Change in validation loss over the training period gives a glance on model convergence. Model convergence graphs show performance of model training and how efficient a model is in converging. These graphs show the lowest loss achieved on validation data, and fluctuations in loss which evaluates a model's stability. The graphs provide an initial basis of comparisons for different models.

\noindent
\textbf{The Dice Similarity Coefficient: } Also known as the Sørensen-Dice coefficient, Dice coefficient is a metric used to quantify the similarity or overlap between two sets or groups. In the context of image segmentation and binary classification tasks, the Dice coefficient is commonly employed to evaluate the similarity between two binary masks or regions of interest (ROIs).

Formally, the Dice Similarity Coefficient (DSC) is defined as:
\begin{equation}
DSC = \frac{2 \times |A \cap B|}{|A| + |B|}
\end{equation}
where:
\begin{align*}
& A \text{ is the first set or binary mask (e.g., the predicted segmentation mask);} \\
& B \text{ is the second set or binary mask (e.g., the ground truth or reference mask);} \\
& |\cdot| \text{ denotes the cardinality of a set, i.e., the number of elements in the set;} \\
& \cap \text{ denotes the intersection operation.}
\end{align*}
The Dice coefficient produces a value between 0 and 1, where:
\begin{itemize}
\item $DSC = 0$ indicates no overlap or dissimilarity between the two sets. It means that there is no commonality between the predicted and reference masks.
\item $DSC = 1$ indicates perfect overlap or similarity between the two sets. It means that the predicted mask perfectly matches the reference mask.
\end{itemize}
In the context of image segmentation, the Dice coefficient is a valuable metric because it measures the agreement between the segmented region and the ground truth. It quantifies how well the segmentation result matches the true region of interest. Higher DSC values indicate better segmentation performance.

\noindent
\textbf{Intersection over Union (IoU) or Jaccard Index: }
IoU measures the overlap between the predicted segmentation mask (\(A\)) and the ground truth mask (\(B\)). It is calculated as the intersection of the two masks divided by their union. A higher IoU indicates better segmentation accuracy.
\[
IoU = \frac{|A \cap B|}{|A \cup B|}
\]
where:
\begin{align*}
& A \text{ is the predicted mask;} \\
& B \text{ is the ground truth mask.}
\end{align*}
In this formula, \(|A \cap B|\) denotes the cardinality of the intersection of sets \(A\) and \(B\), and \(|A \cup B|\) represents the cardinality of their union. IoU quantifies the extent to which the predicted mask and the ground truth mask overlap, providing a valuable measure of segmentation accuracy. The implementation of Jaccard index using python is given below:

\noindent
\textbf{Confusion Matrix: }A confusion matrix provides a detailed breakdown of true positive, true negative, false positive, and false negative predictions. It is useful for understanding the model's performance on different classes or categories within the segmentation task. This is implemented using \verb|confusion_matrix| function from python's \verb|sklearn|.

\noindent
\textbf{Precision: }Precision assesses the accuracy of an algorithm in correctly identifying relevant pixels or regions. It's the ratio of true positive pixels (correctly segmented) to all pixels identified as positive by the algorithm. High precision indicates that when the algorithm marks a pixel or region as part of the target object, it's usually correct. In medical image segmentation, high precision means that when the algorithm identifies an area as a specific organ or structure, it's likely to be accurate, reducing false positives.

\noindent
\textbf{Recall: }Recall, also called sensitivity or true positive rate, gauges an algorithm's capacity to accurately identify all relevant pixels or regions in an image. It's the ratio of true positive pixels to the total pixels constituting the actual target object or region in the ground truth. A high recall value signifies that the algorithm excels at locating and encompassing most of the genuine target object or region. In medical image segmentation, a high recall means the algorithm effectively identifies and includes most relevant anatomical structures, reducing the likelihood of false negatives.

\section{Brain Tumor Segmentation}
The task of Brain tumor segmentation involves the process of identifying and delineating the boundaries of brain tumors in medical images, specifically in brain MRI scans. The goal of this segmentation task is to automatically outline the shape and extent of lower-grade gliomas (LGG) within the brain images.

\subsection{Pre-processing}
The dataset used in this study was obtained from Kaggle \cite{brain_data} and was originally sourced from The Cancer Genome Atlas Low Grade Glioma Collection (TCGA-LGG)\cite{pedano2016tcga}. It includes brain MR images that are accompanied by manually created FLAIR abnormality segmentation masks. The dataset contains MRI FLAIR image data for 110 patients. Each MRI image is an RGB image with three channels, and each mask is a 2D black and white image.

The dataset originally contained 1200 patient images and masks, with 420 masks indicating the presence of tumors. To focus the model on tumor segmentation, images without tumor annotations were removed. The dataset was then split into training, testing, and validation sets using an 8:1:1 ratio. To handle this data efficiently, a Data Generator was employed—a crucial tool in deep learning, particularly for large datasets that don't fit in memory. Data Generators process data in smaller batches during training, effectively managing computational resources and ensuring real-time preprocessing during model training. The pre-processing steps each image goes through are listed below:
\begin{enumerate}
    \item \textbf{Image Resizing}: Images in the dataset were resized to a standard 256 by 256 pixel dimension to ensure compatibility with neural network architectures. This choice balances between preserving important details, which smaller sizes might lose, and avoiding unnecessary noise, which larger sizes could introduce.
        
    \item \textbf{Standardization}: Image and mask data were standardized by adjusting their pixel values to have a mean of 0 and a standard deviation of 1. This uniform scaling simplifies data for deep learning models, promoting convergence and training stability.
        
    \item \textbf{Normalisation of mask images}: The mask images, initially with binary values (0 for background, 1 for the mask), had their values become floating-point during resizing. To prepare them for model training, their dimensions were expanded by one to (256x256x1), followed by a thresholding operation. Pixel values greater than 0 were set to 1 (indicating a tumor), while values equal to or less than 0 were set to 0 (representing the background), maintaining binary suitability for training.
\end{enumerate}
\subsection{Model Training}
\subsubsection{Loss Function}
\textbf{Binary Focal Loss} is a specialized loss function used in binary classification tasks, particularly when dealing with imbalanced datasets or cases where certain classes are of greater interest than others. It is designed to address the problem of class imbalance and focuses on improving the learning of the minority class.
Formally, the Binary Focal Loss (BFL) is defined as follows:
\begin{equation}
BFL = - (1 - p_t)^\gamma \cdot \log(p_t)
\end{equation}
where:
\begin{align*}
& p_t \text{ represents the predicted probability of the true class label;} \\
& \gamma \text{ is a tunable hyperparameter known as the focusing parameter;} \\
& \log(\cdot) \text{ is the natural logarithm.}
\end{align*}
The Binary Focal Loss has the following key characteristics:
\begin{itemize}
\item It introduces the focusing parameter $\gamma$ to control the degree of importance assigned to different examples. A higher $\gamma$ values emphasize the training on hard, misclassified examples, while lower values make the loss less sensitive to those examples.
\item When $\gamma = 0$, the Binary Focal Loss reduces to the standard binary cross-entropy loss.
\item The term $(1 - p_t)^\gamma$ is a modulating factor that reduces the loss for well-classified examples ($p_t$ close to 1) and increases the loss for misclassified examples ($p_t$ close to 0).
\item BFL helps the model focus more on the minority class, which is especially useful in imbalanced datasets where the majority class dominates.
\item It encourages the model to learn better representations for challenging examples, potentially improving overall classification performance.
\item The loss is applied independently to each example in a batch of data during training.
\end{itemize}
The Binary Focal Loss is a valuable tool in addressing class imbalance and improving the training of models for imbalanced binary classification tasks. By introducing the focusing parameter, it allows practitioners to fine-tune the loss function according to the specific characteristics of their dataset and the importance of different classes.
\subsubsection{Model Design Choices}
\textbf{UNet: }The UNet model has input shape of (256, 256,3) for the rgb images and an output layer of shape (256, 256, 1) for the mask output. The final output layer consists of 1x1 convolutional layers followed by batch normalization and sigmoid activation. These layers produce the segmentation mask, where each pixel is classified as either part of the object or background. Sigmoid activation is used for binary segmentation. The model has a total of 31,402,501 parameters with 31,390,723 being trainable.

\noindent
\textbf{Res-UNet: }The Res-UNet model takes RGB images with an input shape of (256, 256, 3) and produces a mask output with an output layer of shape (256, 256, 1). The last layer of the model comprises 1x1 convolutional layers, which are followed by batch normalization and a sigmoid activation function.

\noindent
\textbf{Attention Res-UNet: }Attention Res-UNet follows the same input and output configuration as the previous models due to the same input and output image and mask specifications. 
\subsubsection{Callbacks}
Three callbacks are assigned to the models:
\begin{enumerate}
    \item EarlyStopping Callback (EarlyStopping): The EarlyStopping callback monitors validation loss during training. If there's no improvement (decrease) for 20 epochs, training stops early to prevent overfitting and save time, ensuring the model doesn't learn noise or deviate from the optimal solution.
    \item ReduceLROnPlateau Callback: The ReduceLROnPlateau callback is used to optimize the model's training by lowering the learning rate when the validation loss reaches a plateau or stops improving, aiding the model in fine-tuning and avoiding local minima. The callback monitors 'val\_loss' with a 'min' mode, providing informative updates (verbose=1). It adjusts the learning rate if there's no improvement in validation loss for 10 consecutive epochs (patience=10) by a factor of 0.2 (reduced to 20\% of its previous value), ensuring meaningful improvements with a 'min\_delta' parameter set at 0.0001.
    \item Checkpointer: Checkpointers are specified which saves weights of the trained model only when the validation loss improves. 
\end{enumerate}
\subsubsection{Model Compilation and Fitting}
The models are compiled using the Adam optimizer with an initial learning rate of '1e-5.' Multiple initial learning rates were tested, with higher rates causing divergence and lower rates slowing down training. Two compilations are done for each model, one using the dice-coefficient as the loss function and the other using Binary Focal loss. Training is performed on both the training and validation data for 100 epochs initially.
\subsection{Results}
\begin{table}
    \centering
    \begin{tabular}{|c|c|c|c|}
        \hline
        Models & Execution Time & Epoch\\
        \hline
        UNet & 34 min 20 sec & 69\\
        \hline
        Res-UNet & 42 min 1 sec & 89\\
        \hline
        Attention Res-UNet & 1 hr 1 min & 100\\
        \hline
    \end{tabular}
    \caption{Execution time and epochs of trained models}
    \label{tab: brain_exec}
\end{table}
The model training times and epochs run are listed in Table \ref{tab: brain_exec}.
The differences in execution times among the models indicate varying computational resource requirements during training. Notably, Attention Res-UNet emerges as the model with the longest training duration. This extended duration could be attributed to the model's complexity that necessitated additional time for convergence.

Regarding training behavior, the number of epochs completed by each model offers insights into their respective convergence behaviors. The UNet model exhibits a comparatively lower number of epochs, implying a relatively swift convergence. This is indicative of a particularly efficient training process. Conversely, Res-UNet and Attention Res-UNet underwent more extensive training, implying potentially more intricate model architectures or the need for extended training periods to achieve convergence. 

Furthermore, it's worth noting that some models concluded training prematurely due to a lack of improvement in validation loss, as evidenced by their lower epoch counts. This highlights the consideration of early stopping strategies, a common technique used to curtail training and prevent overfitting. This observation raises the need for discussions on optimizing model performance and making thoughtful decisions about resource allocation during training.
\begin{figure}[h]
    \centering
    \begin{subfigure}[b]{0.3\textwidth}
        \includegraphics[width=\textwidth]{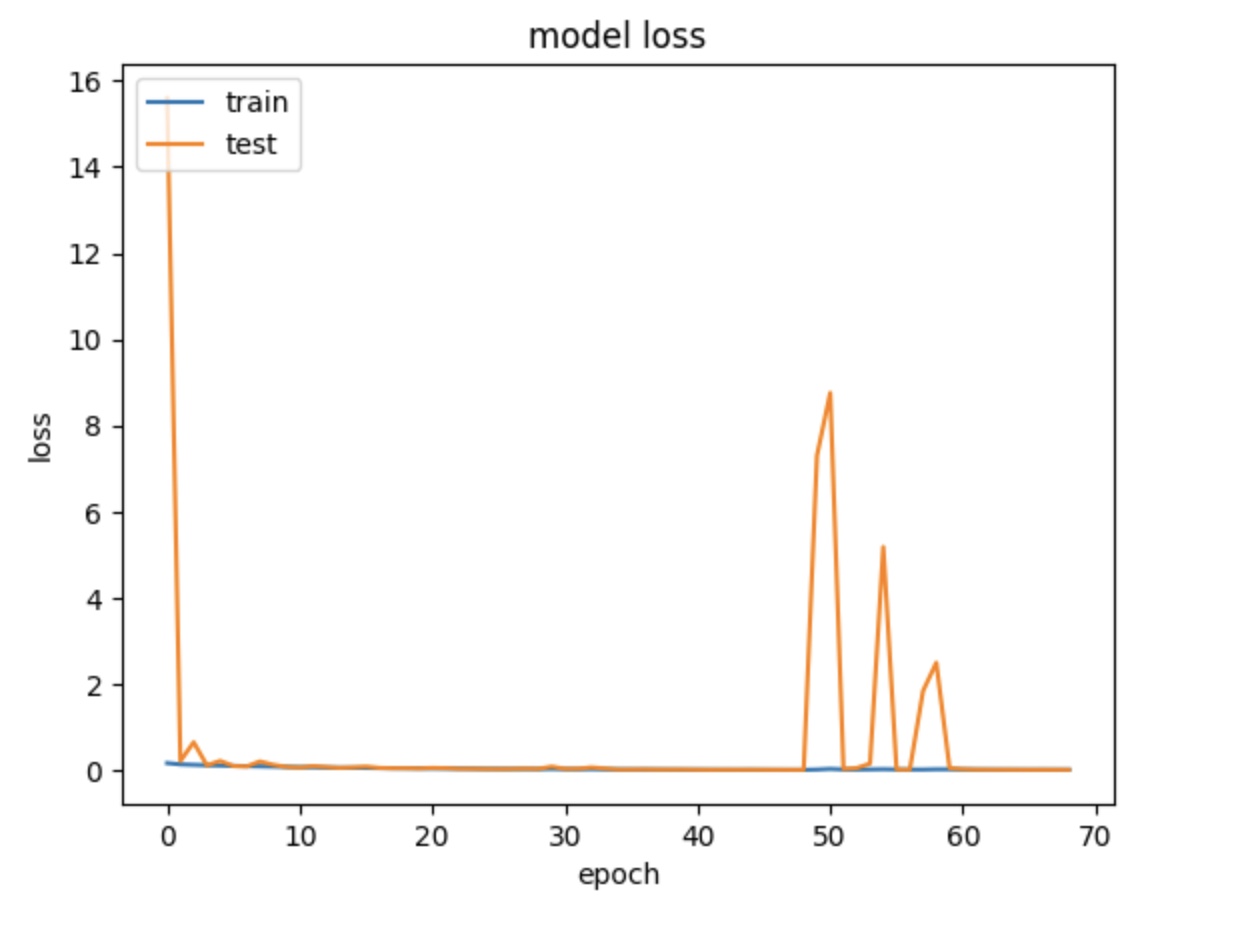}
        \caption{UNet}
    \end{subfigure}
    \begin{subfigure}[b]{0.3\textwidth}
        \includegraphics[width=\textwidth]{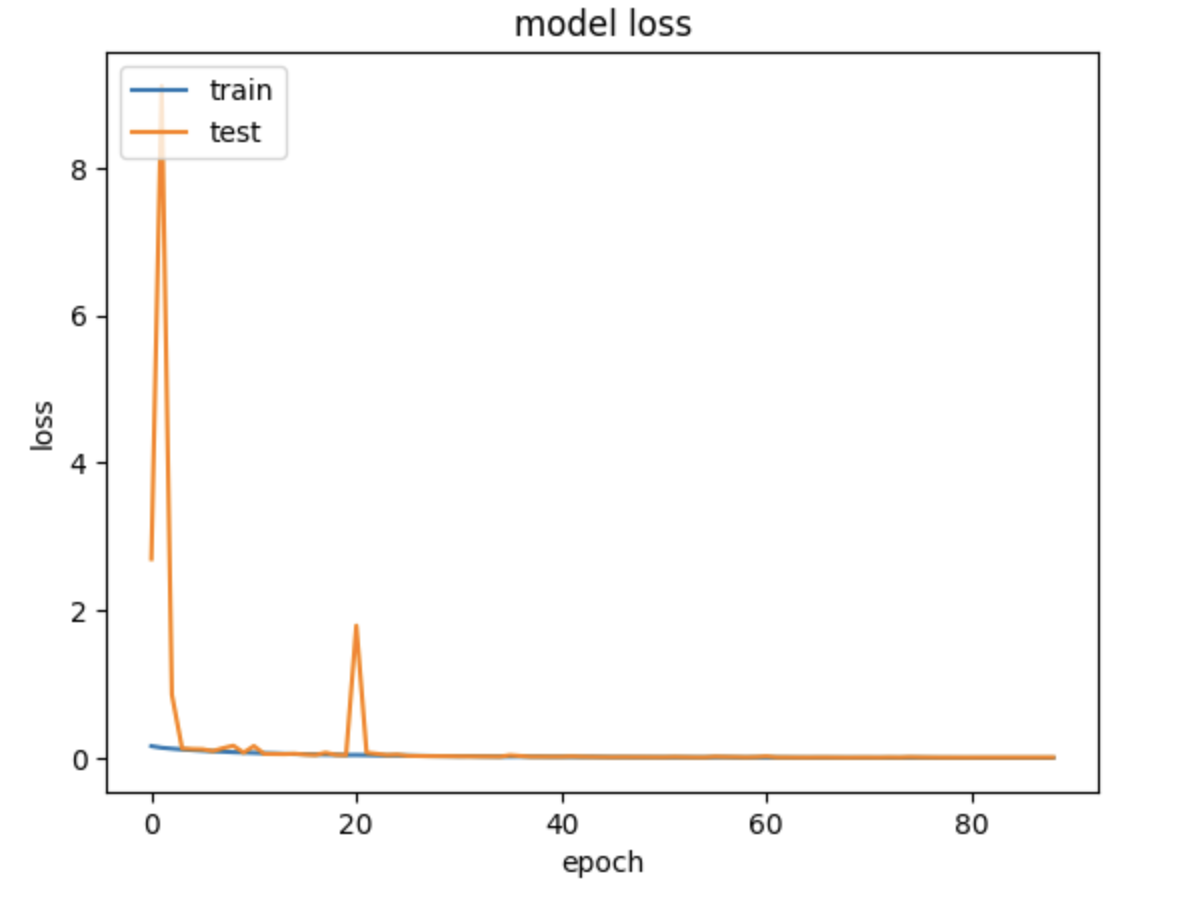}
        \caption{Res-UNet}
    \end{subfigure}
    \begin{subfigure}[b]{0.3\textwidth}
        \includegraphics[width=\textwidth]{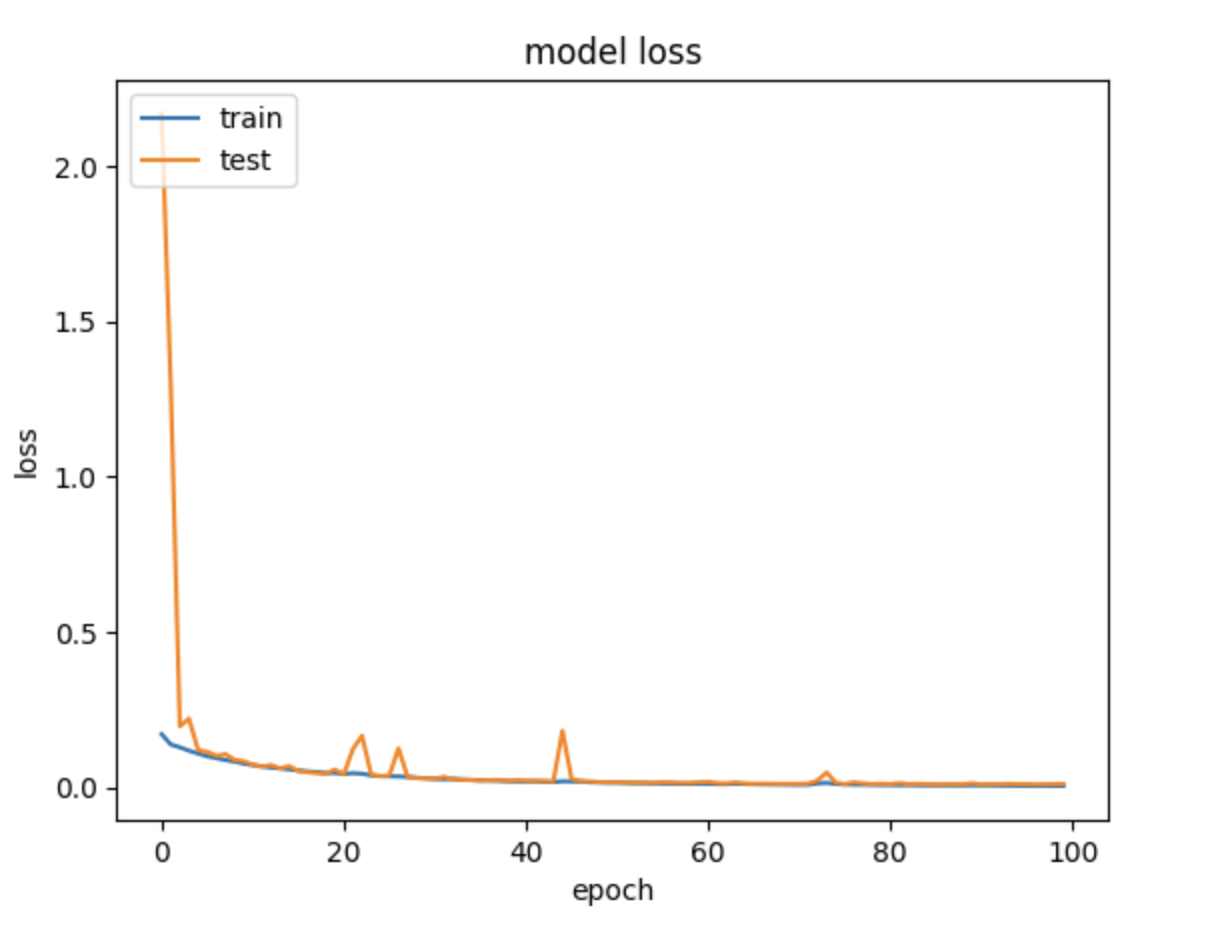}
        \caption{Attention Res-UNet}
    \end{subfigure}
    \caption{Change in Binary focal loss for each model}
    \label{fig: model_convergence_brain}
\end{figure}
The sub-figures in Figure \ref{fig: model_convergence_brain} depict the evolution of the Binary Focal Loss over epochs for training and validation data for three different models. These results offer insights into how these models perform during the training process.
\begin{enumerate}
    \item \textbf{Initial Validation Loss:} Initial validation losses vary among the models. UNet starts with a high initial loss (around 15), indicating initial difficulty in accurate predictions. In contrast, Res-UNet begins with a lower loss (around 8), while Attention Res-UNet starts with an even lower loss (approximately 2), suggesting that the latter two models make relatively better predictions from the start.
    
    \item \textbf{Early Epoch Performance:} All three models exhibit a rapid decrease in validation loss within the first ten epochs. This implies that they quickly learn to capture relevant patterns in the data and improve their predictions during this early training phase.
    
    \item \textbf{Stability in Training:} During training, all models maintain generally low validation losses, with some fluctuations. UNet exhibits significant fluctuations towards training's end, suggesting sensitivity to data variations. In contrast, Res-UNet shows minor early fluctuations but stabilizes. Attention Res-UNet also experiences initial fluctuations, but they are much smaller than in the other models.
    
    \item \textbf{Comparison of Model Performance:} UNet quickly reduces validation loss at the start but has higher fluctuations later. Res-UNet starts with a moderate loss, has some early fluctuations, and stabilizes. Attention Res-UNet consistently performs well from the beginning with minimal fluctuations.
\end{enumerate}
Overall, these results highlight trade-offs between rapid initial learning and stability in model performance. UNet learns quickly but exhibits greater instability, while Res-UNet and Attention Res-UNet provide more consistent and reliable predictions.
Table \ref{tab:brain_model_performance} provides performance metrics for UNet, Res-UNet, and Attention Res-UNet, when applied to test data.
\begin{table}[ht]
    \centering
    \caption{Performance Metrics for UNet, Res-UNet, and Attention Res-UNet on test data}
    \begin{tabular}{lcccccc}
        \hline
        Model & Focal Loss & Accuracy & Precision & Recall & Dice & IoU \\
        \hline
        UNet & 0.0169 & 0.987 & 0.852 & 0.623 & 0.72 & 0.563\\
        Res-UNet & 0.0062 & \textbf{0.996} & \textbf{0.923} & 0.939 & \textbf{0.931} & \textbf{0.870} \\
        Attention Res-UNet & \textbf{0.0055} & \textbf{0.996} & 0.902 & \textbf{0.946} & 0.923 & 0.858 \\
        \hline
    \end{tabular}
    \label{tab:brain_model_performance}
\end{table}

\noindent
\textbf{1. Focal Loss:} All the models achieve low focal loss with Res-UNet and Attention Res-UNet outperforming UNet. Attention Res-UNet achieves the lowest Focal Loss, highlighting its proficiency in addressing class imbalance. This means that the variants perform better at focusing on hard to classify pixels, which, in this case, is the tumor class.

\noindent
\textbf{2. Accuracy:} Res-UNet and Attention Res-UNet exhibit impressive accuracies, approximately 99.6\%, surpassing UNet, which achieves 98.7\%. Both Res-UNet and Attention Res-UNet excel in pixel-level classification accuracy.

\noindent
\textbf{3. Precision and Recall:} Res-UNet demonstrates superior precision, indicating accurate positive pixel classification with minimal false positives. UNet and Attention Res-UNet exhibit slightly lower precision values. Conversely, Attention Res-UNet achieves the highest recall, suggesting its effectiveness in capturing a larger proportion of true positives.

\noindent
\textbf{4. Dice Coefficient:} Res-UNet achieves the highest Dice coefficient at approximately 0.931, signifying accurate spatial predictions. UNet and Attention Res-UNet yield slightly lower Dice coefficients but maintain strong performance.

\noindent
\textbf{5. Intersection over Union (IoU):} Res-UNet achieves the highest IoU of approximately 0.870, indicating superior spatial overlap. UNet and Attention Res-UNet record slightly lower IoU values, though they continue to deliver commendable results in this aspect.

In summary, Res-UNet and Attention Res-UNet consistently outperform UNet across multiple performance metrics, underscoring their superior performance in image segmentation on the test data. Res-UNet excels in precision, Dice coefficient, and IoU, while Attention Res-UNet achieves the highest recall.

\subsection{Discussions}
\begin{figure}[h!]
  \centering
  \subfloat{\includegraphics[width=0.8\textwidth]{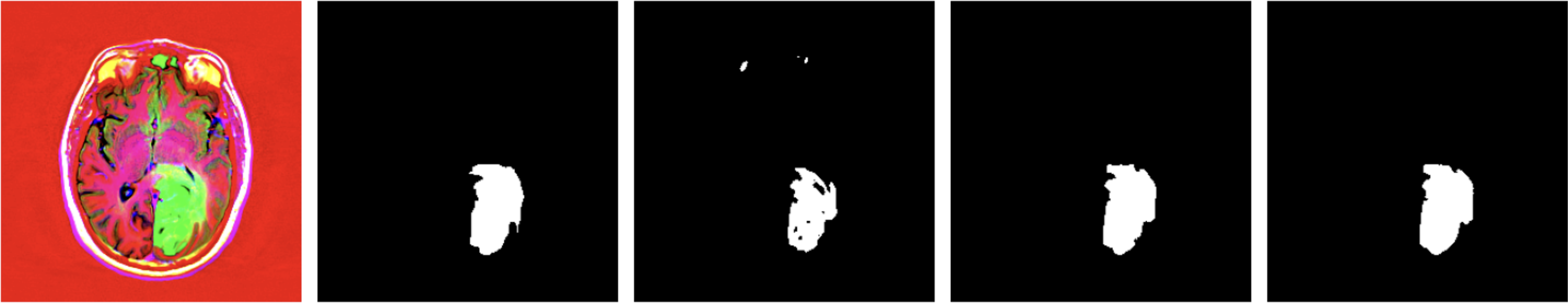}}\\
  \subfloat{\includegraphics[width=0.8\textwidth]{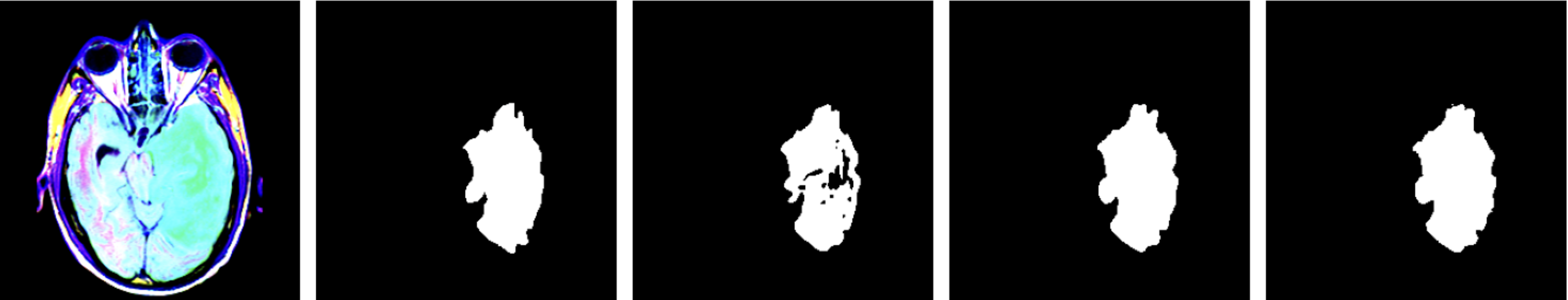}}\\
  \subfloat{\includegraphics[width=0.8\textwidth]{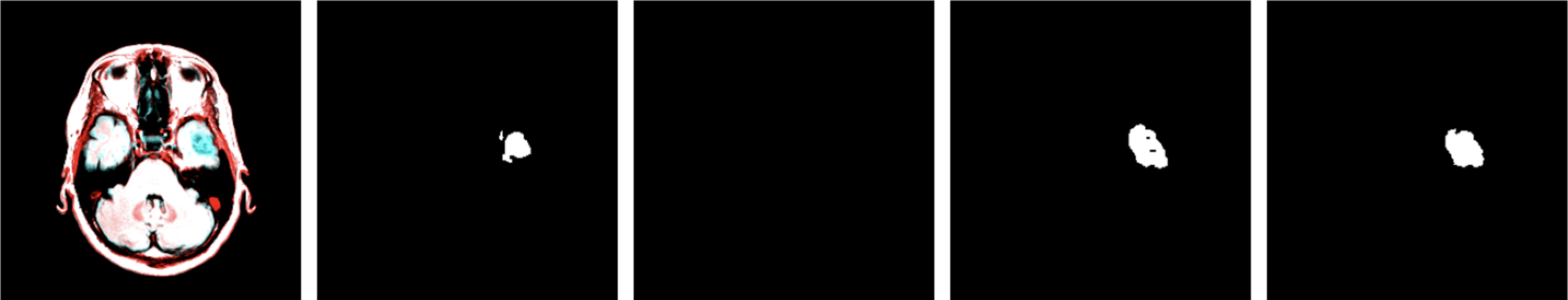}}\\
  \subfloat{\includegraphics[width=0.8\textwidth]{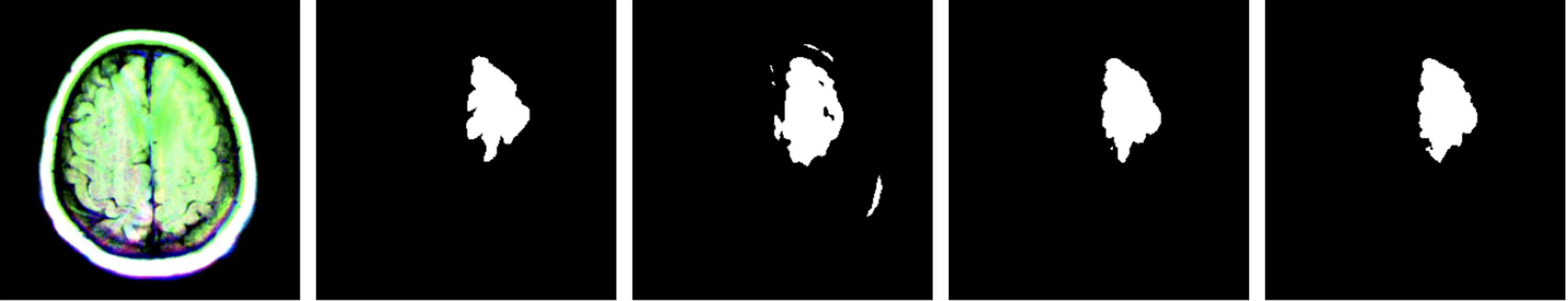}}
  \caption{Segmentation results by the three models for four different examples, from left to right are the input images, ground-truth, segmentation results by UNet, Res-UNet and Attention Res-UNet.}
  \label{fig: brain_models_res}
\end{figure}
Figure \ref{fig: brain_models_res} shows four examples with given image and ground truth mask followed my predictions from the three models. The four examples were chosen as they represent the different types of results observed in the whole test prediction.

UNet exhibits sensitivity to tumor features and shows promise in identifying likely tumor locations but tends to misclassify tumor pixels as background, leading to false negatives. It also mistakenly classifies background pixels as tumors, causing false positives and impacting precision. Res-UNet and Attention Res-UNet, on the other hand, deliver highly accurate predictions, capturing fine details and maintaining a balance between sensitivity and specificity. While they occasionally overestimate tumor presence, these misclassifications are minor.

UNet and its variants perform adequately in most cases but struggle when tumors are very small or have complex boundaries. They also face challenges with class imbalance, resulting in misclassification and poor recall. Res-UNet and Attention Res-UNet mitigate these limitations, successfully locating tumors in challenging conditions and reducing misclassifications significantly. Attention Res-UNet excels in handling class imbalance.

Despite variations in performance, all models achieve high accuracy scores. However, accuracy may not be a reliable metric as it can remain high even when tumors are misclassified due to the relatively small size of tumors compared to background, making it unreliable for assessing model performance.
\section{Polyp Segmentation}
Polyp segmentation refers to the process of identifying and delineating the boundaries of polyps in medical images, particularly in the context of medical imaging, endoscopy, and colonoscopy. The goal of this segmentation task is to automatically outline the shape and extent of polyps from colonoscopy images.


\subsection{Pre-processing}
The CVC-ClinicDB dataset \cite{Polyp_dataset} is utilized for the segmentation task, featuring frames extracted from colonoscopy videos showcasing polyps. It includes corresponding ground truth masks outlining polyp regions. The dataset consists of two main types of images: original colonoscopy frames accessible at 'original/frame\_number.tiff' and corresponding polyp masks at 'ground truth/frame\_number.tiff'.

A Pandas DataFrame is employed to manage image and mask paths. The DataFrame is used to split the data into training, testing, and validation sets in an 8:1:1 ratio. A Dataset generator processes images and masks in the training and validation data one by one, using a 'tf\_parse()' function to read, resize, and preprocess them for compatibility with the program's requirements. The pre-processing steps are listed below:
\begin{enumerate}
    \item \textbf{Reading the Image:} The function first reads the image from the file path \(x\) using OpenCV (\texttt{cv2.imread}). This reads the image as it is in its original form.
    \item \textbf{Resizing the Image:} After reading, the image is resized to a fixed size of \(256\times256\) pixels using OpenCV's \texttt{cv2.resize} function. This resizing ensures that all images have the same dimensions, which is typically necessary for training deep learning models.
    \item \textbf{Normalizing the Image:} The pixel values of the resized image are scaled to a range between 0 and 1. This is done by dividing all pixel values by 255.0. Normalizing the pixel values helps the deep learning model learn more effectively.
\end{enumerate}
\subsection{Model Training}
\subsubsection{Loss Function}
Binary Focal Loss is used as the loss function for the models. The masks in the problem are binary(label and background) and thus follows a similar design as the Brain Tumor Problem. 
\subsubsection{Model Design Choices}
The model design choices for \textbf{UNet}, \textbf{Res-UNet} and \textbf{Attention Res-UNet} for Polyp segmentation are similar to the the models used for Brain Tumor Segmentation. The two problems, even though crucial in their own ways to the medical community, shares the same configuration in the fact that they involve creating binary segmentation masks from RGB images. Hence the input shape for the images in both the problems is (256, 256, 3) while the output shape is (256, 256, 1), and this does not require a change in the model architectures. 
\subsubsection{Callbacks}
The callbacks used for this problem are Early Stopping, Reducing Learning rate and checkpointers as mentioned in the previous problem. 
\subsubsection{Model Compiling and Fitting}
Models are compiled with the Adam optimizer with an initial learning rate if \verb|1e-5| and fitted for 100 epochs. 
\subsection{Results}
\begin{table}
    \centering
    \begin{tabular}{|c|c|c|c|}
        \hline
        Models & Execution Time & Epoch\\
        \hline
        UNet & 51 min 14 sec & 73\\
        \hline
        Res-UNet & 45 min 12 sec & 63\\
        \hline
        Attention Res-UNet & 56 min 45 sec & 62\\
        \hline
    \end{tabular}
    \caption{Execution time and epochs of trained models for Polyp Segmentation}
    \label{tab: polyp_exec}
\end{table}
Model training times and epochs run are listed in Table \ref{tab: polyp_exec}.

Attention Res-UNet had the longest training duration, taking 56 minutes and 45 seconds, likely due to its complex architecture. UNet showed efficient training, completing in 73 epochs, while Res-UNet and Attention Res-UNet required more extensive training, with 45 minutes and 12 seconds and 62 epochs for Res-UNet, and the same time and epochs for Attention Res-UNet. Some models ended training early due to no improvement in validation loss, highlighting the importance of early stopping strategies to prevent overfitting. This emphasizes the need for optimizing model performance and resource allocation decisions during training.



\begin{figure}[h]
    \centering
    \includegraphics[width=0.6\textwidth]{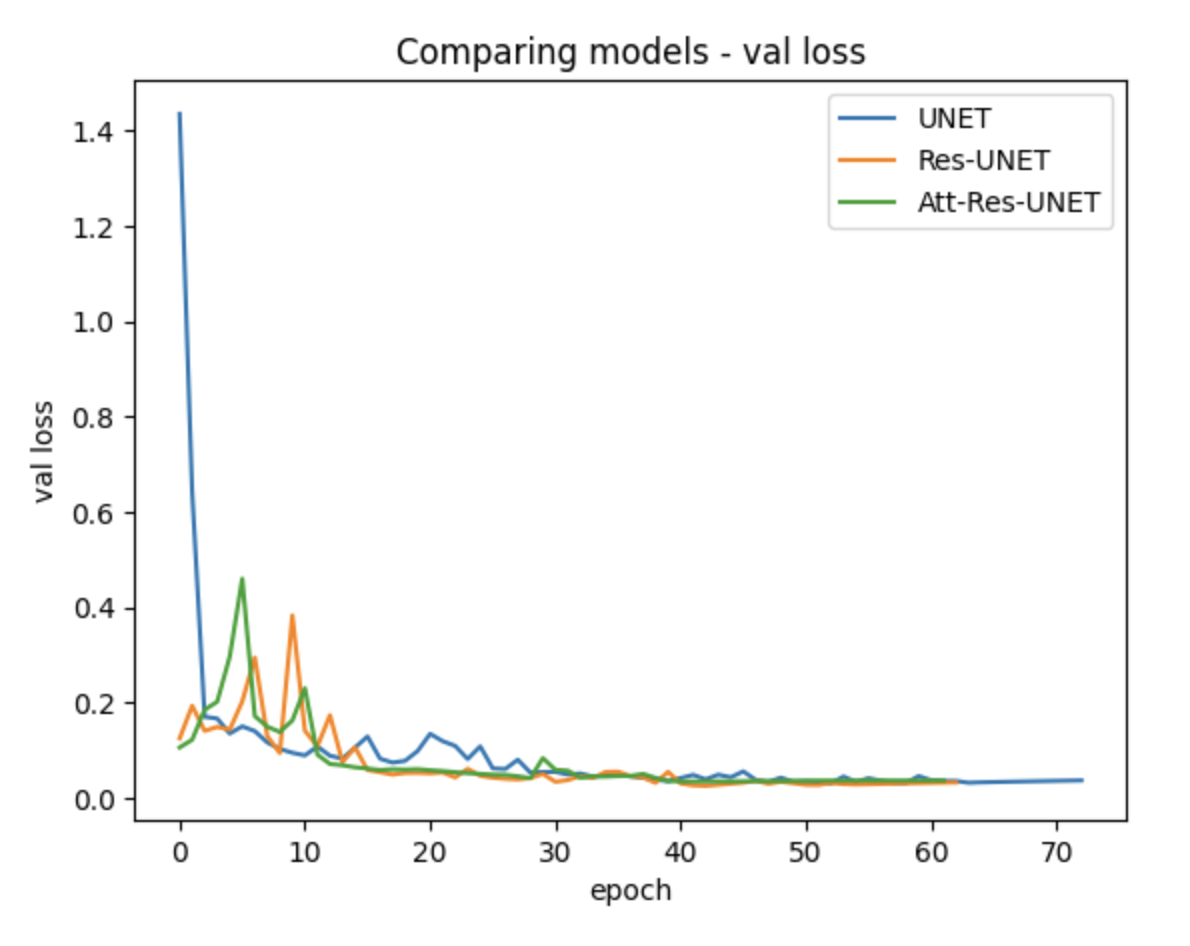}
    \caption{Convergence for trained models on Polyp Segmentation}
    \label{fig: model_convergence_polyp}
\end{figure}
Figure \ref{fig: model_convergence_polyp} depict the evolution of the Binary Focal Loss over epochs for validation data for three different models. These results offer insights into how these models perform during the training process.

\noindent
\textbf{UNet Model:}

\noindent
The UNet model initiates training with a high validation loss, approximately 1.4, primarily because its initial weights are far from optimal. However, within the initial ten epochs, it experiences a rapid decrease in validation loss, a common pattern during the early training stages of many neural networks. This reduction reflects the model's improvement in fitting the training data as it adjusts its weights through techniques like backpropagation and stochastic gradient descent (SGD). Following this initial phase, the UNet model maintains a relatively low and stable loss for the remaining epochs, albeit with some minor fluctuations. These fluctuations are likely attributable to inherent data noise and the stochastic nature of the optimization process.
  
  

\noindent
\textbf{Res-UNet and Attention Res-UNet Models:}

\noindent
Both Res-UNet and Attention Res-UNet models begin training with a low initial validation loss, roughly 0.2, suggesting potential pretraining or initialization, placing them closer to a reasonable starting point. In the initial 15 epochs, both models experience fluctuations in loss, common during initial training phases as they adapt to data and fine-tune weights, possibly indicating sensitivity to initial configuration or data noise. As training progresses, both models achieve stable loss values, signifying they have reached a consistent and relatively optimal solution compared to UNet within this timeframe. Eventually, all models reach a minimum loss of approximately 10\%, demonstrating similar performance levels in minimizing loss on the validation data, despite differences in convergence speed and early fluctuations.

In summary, these results indicate that UNet initiates training with a higher loss but converges swiftly. In contrast, Res-UNet and Attention Res-UNet begin with lower losses but may show more early training fluctuations. Nevertheless, all models ultimately achieve a similar minimum loss, showcasing their capability to capture crucial data features and make accurate predictions.
\begin{table}[ht]
    \centering
    \caption{Performance Metrics for UNet, Res-UNet, and Attention Res-UNet on test data}
    \begin{tabular}{lcccccc}
        \hline
        Model & Focal Loss & Accuracy & Precision & Recall & Dice & IoU \\
        \hline
        UNet & 0.0387 & 0.968 & 0.913 & 0.733 & 0.813 & 0.686\\
        Res-UNet & \textbf{0.0369} & \textbf{0.971} & \textbf{0.925} & 0.766 & \textbf{0.838} & \textbf{0.721} \\
        Attention Res-UNet & 0.0394 & 0.969 & 0.881 & \textbf{0.788} & 0.832 & 0.712 \\
        \hline
    \end{tabular}
    \label{tab:polyp_model_performance}
\end{table}
Table \ref{tab:polyp_model_performance} provides performance metrics for UNet, Res-UNet, and Attention Res-UNet when applied to test data:

\noindent
\textbf{1. Focal Loss:} All the models achieve low Focal Loss values, with Res-UNet and Attention Res-UNet outperforming UNet. Res-UNet achieves the lowest Focal Loss, highlighting its proficiency in addressing class imbalance. This means that the variants perform better at focusing on hard-to-classify pixels, which, in this case, is the tumor class.

\noindent
\textbf{2. Accuracy:} Res-UNet and Attention Res-UNet exhibit impressive accuracies, approximately 99.6\%, surpassing UNet, which achieves 98.7\%. Both Res-UNet and Attention Res-UNet excel in pixel-level classification accuracy.

\noindent
\textbf{3. Precision and Recall:} Res-UNet demonstrates superior precision, indicating accurate positive pixel classification with minimal false positives. UNet and Attention Res-UNet exhibit slightly lower precision values. Conversely, Attention Res-UNet achieves the highest recall, suggesting its effectiveness in capturing a larger proportion of true positives.

\noindent
\textbf{4. Dice Coefficient:} Res-UNet achieves the highest Dice coefficient at approximately 0.931, signifying accurate spatial predictions. UNet and Attention Res-UNet yield slightly lower Dice coefficients but maintain strong performance.

\noindent
\textbf{5. Intersection over Union (IoU):} Res-UNet achieves the highest IoU of approximately 0.870, indicating superior spatial overlap. UNet and Attention Res-UNet record slightly lower IoU values, though they continue to deliver commendable results in this aspect.

In summary, Res-UNet and Attention Res-UNet consistently outperform UNet across multiple performance metrics, underscoring their superior performance in image segmentation on the test data. Res-UNet excels in precision, Dice coefficient, and IoU, while Attention Res-UNet achieves the highest recall.
\begin{figure}[h!]
  \centering
  \subfloat{\includegraphics[width=0.8\textwidth]{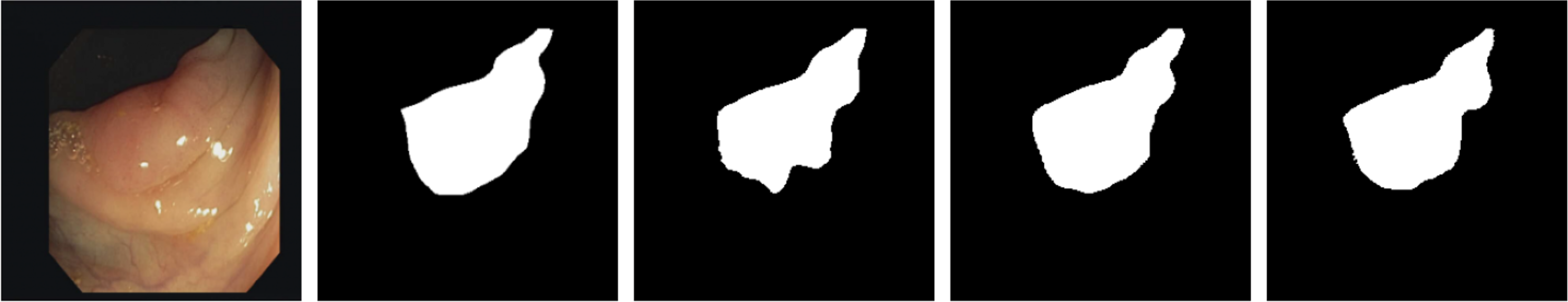}}\\
  \subfloat{\includegraphics[width=0.8\textwidth]{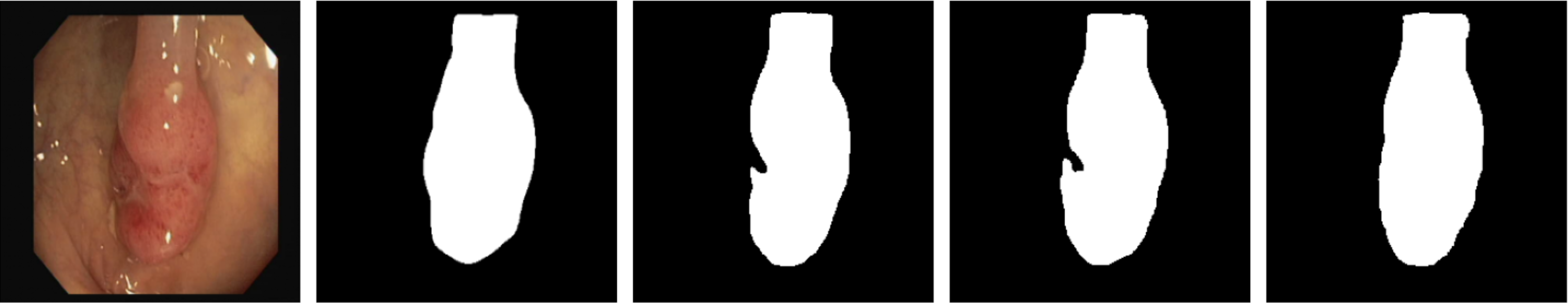}}\\
  \subfloat{\includegraphics[width=0.8\textwidth]{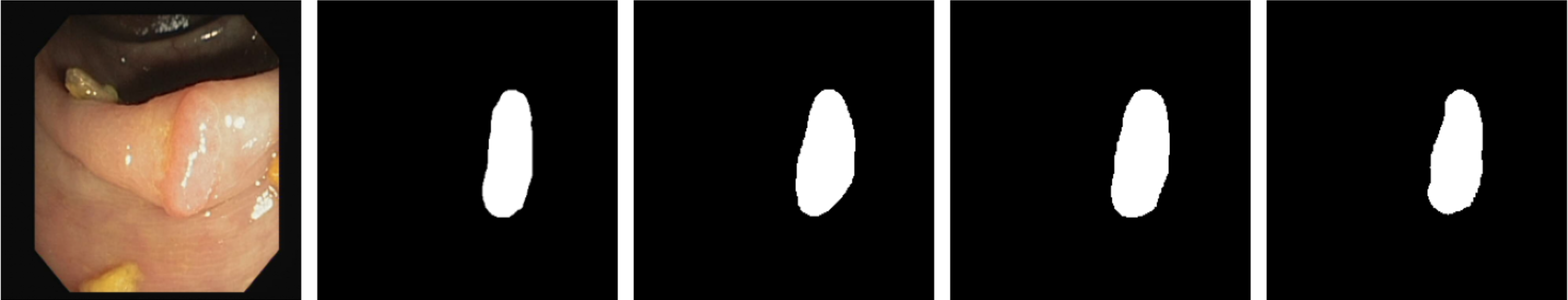}}\\
  \subfloat{\includegraphics[width=0.8\textwidth]{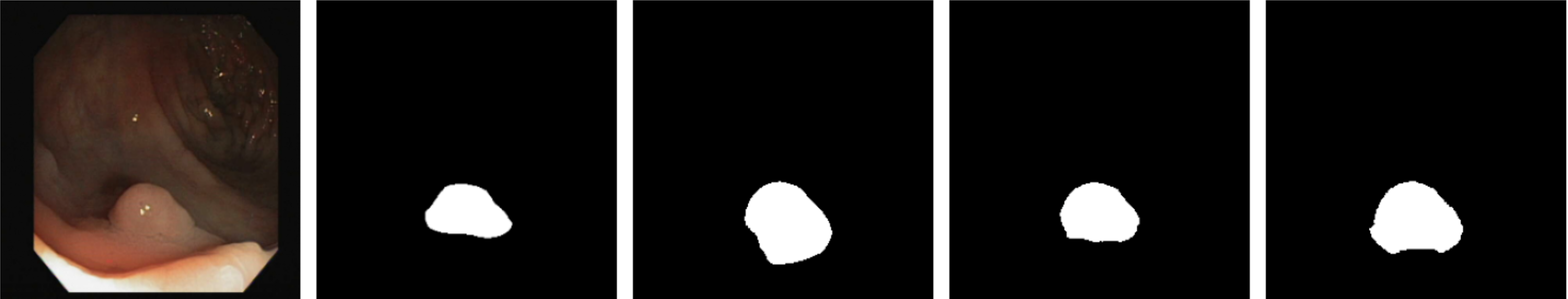}}
  \caption{Segmentation results by the three models for four different examples, from left to right are the input images, ground-truth, segmentation results by UNet, Res-UNet and Attention Res-UNet.}
  \label{fig: polyp_models_res}
\end{figure}
Figure \ref{fig: polyp_models_res} shows four examples with given image and ground truth mask followed my predictions from the three models.
\subsubsection{Discussion}
Polyp segmentation presents challenges due to the irregular and random sizing of polyps, limiting generalization, exacerbated by data limitations. All models trained show above-average segmentation results. UNet, as the base model, trains and converges quickly, especially benefiting from the less imbalanced nature of polyp scans compared to brain MRI masks.

However, UNet exhibits lower performance in predicting the target class, reflecting its sensitivity to polyp features but struggles with class imbalance. It occasionally misclassifies some polyp pixels as background (false negatives) and background pixels as polyps (false positives), impacting both sensitivity and precision. The low True Positive score in the confusion matrix underscores these challenges in accurate polyp detection.

In contrast, Res-UNet and Attention Res-UNet perform consistently, reflecting their performance in brain tumor segmentation. They excel in capturing intricate edge boundaries and maintain accuracy with small ground truth masks. There are rare instances of slight overestimation of polyp presence, but these misclassifications are minor and have minimal impact. Attention Res-UNet is better at predicting true positives than other models reflected by its low recall score.
\section{Heart Segmentation}
The third task involves the multi-label segmentation of cardiac structures in medical images, specifically targeting the Left Ventricle (LV), Right Ventricle (RV), and Myocardium. Accurate segmentation of the LV is essential for assessing its size and function, while RV segmentation aids in diagnosing cardiac conditions. Furthermore, precise Myocardium segmentation provides insights into its thickness and function, offering indicators of heart health and potential issues.
\subsection{Data Pre-processing}
The "Automatic Cardiac Diagnosis Challenge" (ACDC)\cite{ACDC_dataset} dataset is used for this segmentation task. The dataset encompasses data from 150 CMRI recordings which are stored in a 4D "nifti" format, preserving the original image resolution and primarily containing whole short-axis slices of the heart specifying the diastolic and systolic phases of the cardiac cycle. The MRI images are in grayscale, while the mask images employ a 0 to 3 scale, with 0 representing the background, 1 corresponding to the RV cavity, 2 representing the myocardium, and 3 corresponding to the LV cavity.

The preprocessing steps involved creating a dataframe to record image and mask volumes, reading them using the `nibabel` library, and iterating through slices in the third dimension of both the image and mask volumes. Each slice was cropped using a custom `crop` function, with most images having a minimum dimension less than 150, leading to a final size of (128, 128) to avoid introducing noise or unreliable information.

Mask images, with pixel values ranging from 0 to 3 (representing labels and background), were converted to \textbf{one-hot encoding} by increasing the dimensionality to 4, a crucial step for multi-label loss functions and more accurate predictions. For instance, a pixel value of 0 became (0, 0, 0, 0), and 3 became (0, 0, 0, 1).

MRI pixel values, with a maximum of 3049, were \textbf{normalized} to a range of 0 to 1, making them compatible with neural networks. These preprocessing steps were essential for preparing the data for model training.
\subsection{Model Training}
\subsubsection{Loss Function}
Categorical Focal Loss is used as the loss function for the multi label segmentation task.

\noindent
\textbf{Categorical Focal Cross-Entropy} combines the concepts of categorical cross-entropy and focal loss to create a loss function suitable for multi-class segmentation tasks with class imbalance. It introduces the focal loss component into the standard categorical cross-entropy. This helps the model focus on harder-to-classify pixels while handling imbalanced datasets.
\begin{equation}
CFC(y, p) = -\sum_{i=1}^{N} \alpha_i \cdot (1 - p_i)^\gamma \cdot y_i \cdot \log(p_i)
\end{equation}
In summary, Categorical Focal Cross-Entropy is a loss function that blends the properties of categorical cross-entropy and focal loss to improve the training of models on imbalanced multi-class segmentation tasks. It helps the model pay more attention to minority classes and focus on pixels that are difficult to classify. The \verb|CategoricalFocalCrossentropy| loss function is implemented from \verb|keras.losses| library.
\subsubsection{Model Design Choices}
\textbf{Input and output shapes}: The task of multi label segmentation and nature of greyscale MRI images require the output mask shape of the models to have a size of (128, 128, 4) and input shape to be (128, 128, 1). This in turn reduces the total number of parameters in the model when compared to the previous problems.

\noindent
\textbf{Activation Function}: Softmax Classifier is used as the activation function in the output layer of all the models, as it is equipped to run classification/segmentation on multi-labelled prediction.

\noindent
\textbf{UNet}
\begin{figure}[h]
    \centering
    \includegraphics[width=0.8\textwidth]{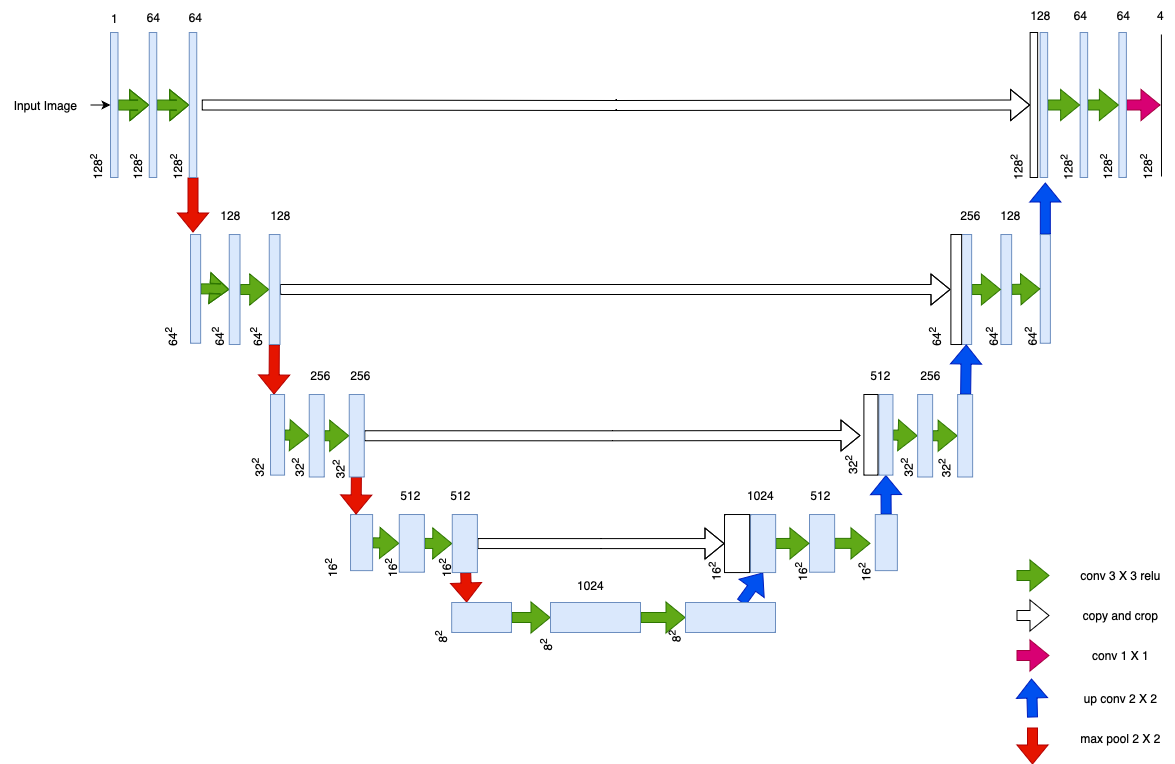}
    \caption{UNet Architecture}
    \label{fig:unet_heart}
\end{figure}
Number of parameters: 31401556

\noindent
\textbf{Res-UNet}
\begin{figure}[h]
    \centering
    \includegraphics[width=0.8\textwidth]{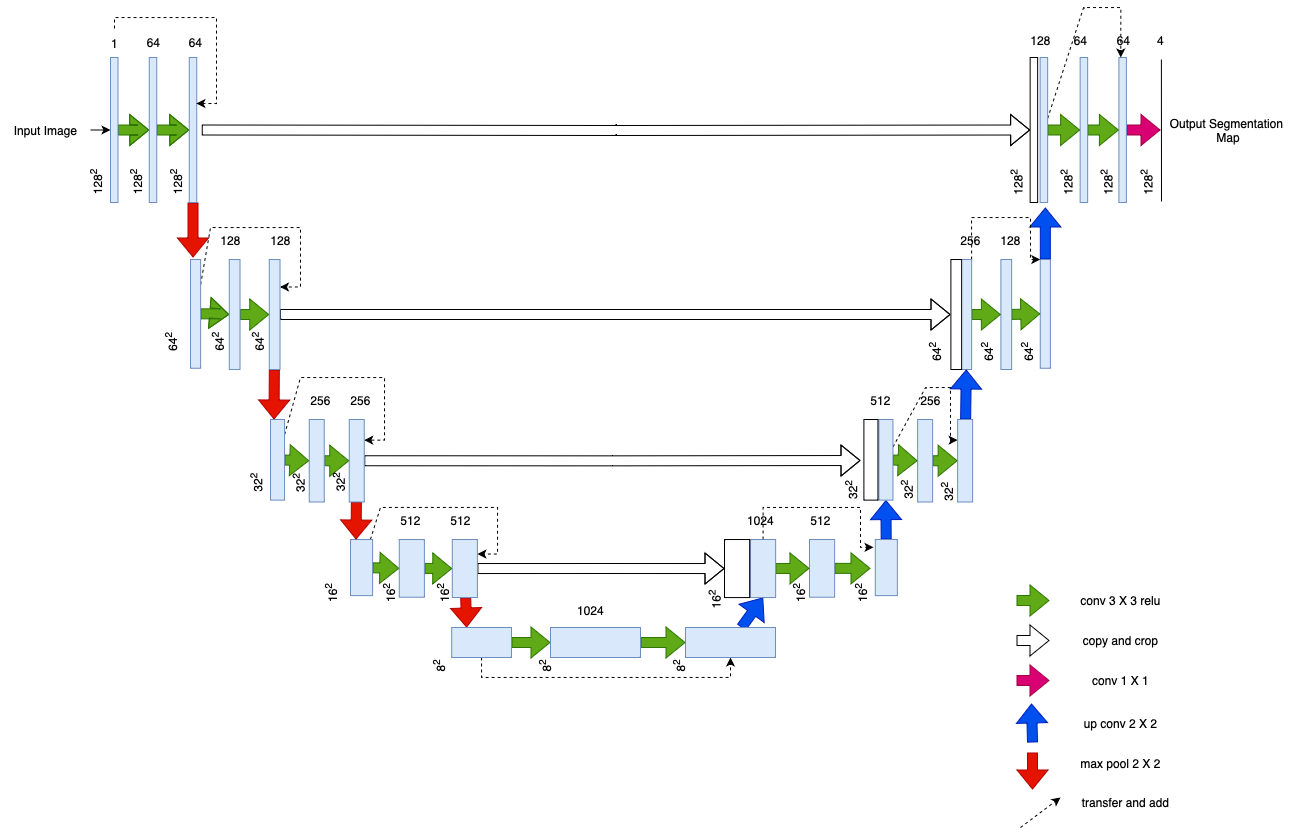}
    \caption{Res-UNet Architecture}
    \label{fig:res_unet_heart}
\end{figure}
Number of parameters: 33157140

\noindent
\textbf{Attention Res-UNet}
\begin{figure}[h]
    \centering
    \includegraphics[width=0.8\textwidth]{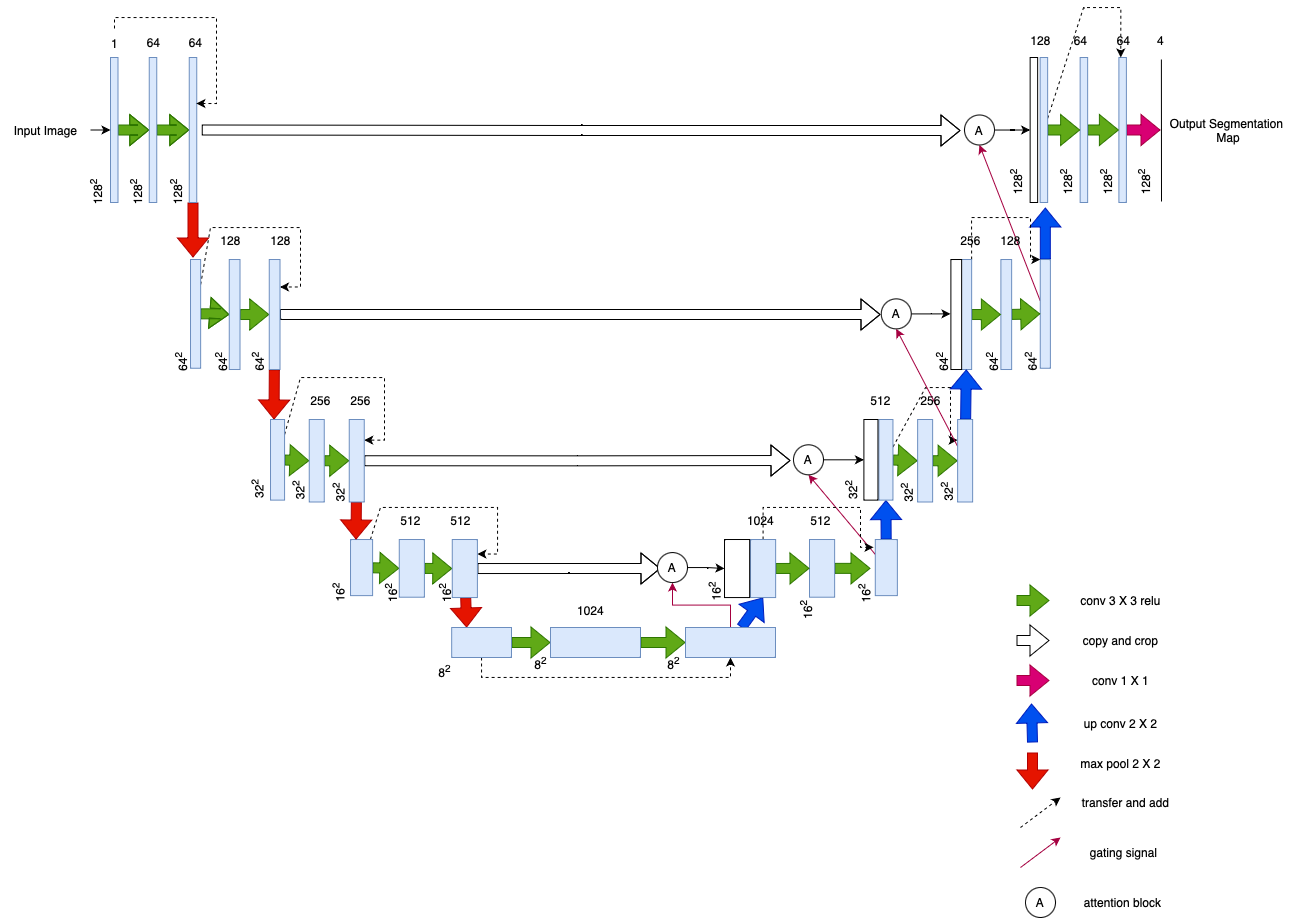}
    \caption{Attention Res-UNet Architecture}
    \label{fig:att_res_unet_heart}
\end{figure}
Number of parameters: 39089304
\subsubsection{Model Compiling and Fitting}
All the models were compiled with Adam optimizer at initial learning rate of 1e-5 and fitted with Early Stopping, Reduce Learning Rate and Checkpointer callbacks for 100 epochs. 
\subsection{Results}
\begin{table}
    \centering
    \begin{tabular}{|c|c|c|c|}
        \hline
        Models & Execution Time & Epoch\\
        \hline
        UNet & 19 min & 86\\
        \hline
        Res-UNet & 25 min 20 sec & 98\\
        \hline
        Attention Res-UNet & 27 min 48 sec & 83\\
        \hline
    \end{tabular}
    \caption{Execution time and epochs of trained models for Multi-label Heart Segmentation}
    \label{tab: heart_exec}
\end{table}
Model training times and epochs run are listed in Table \ref{tab: heart_exec}.

UNet had the shortest training duration, taking 19 minutes, but it required 86 training epochs to reach convergence. In contrast, Res-UNet had a longer training duration, lasting 25 minutes and 20 seconds, and it completed 98 training epochs before converging. Attention Res-UNet, with the longest training duration at 27 minutes and 48 seconds, reached convergence after 83 training epochs.

These results illustrate the trade-offs between training time and the number of epochs required for these models. UNet trained relatively quickly but needed more epochs, while Res-UNet and Attention Res-UNet took more time but required fewer epochs to achieve convergence.

\begin{figure}[h]
    \centering
    \includegraphics[width=0.6\textwidth]{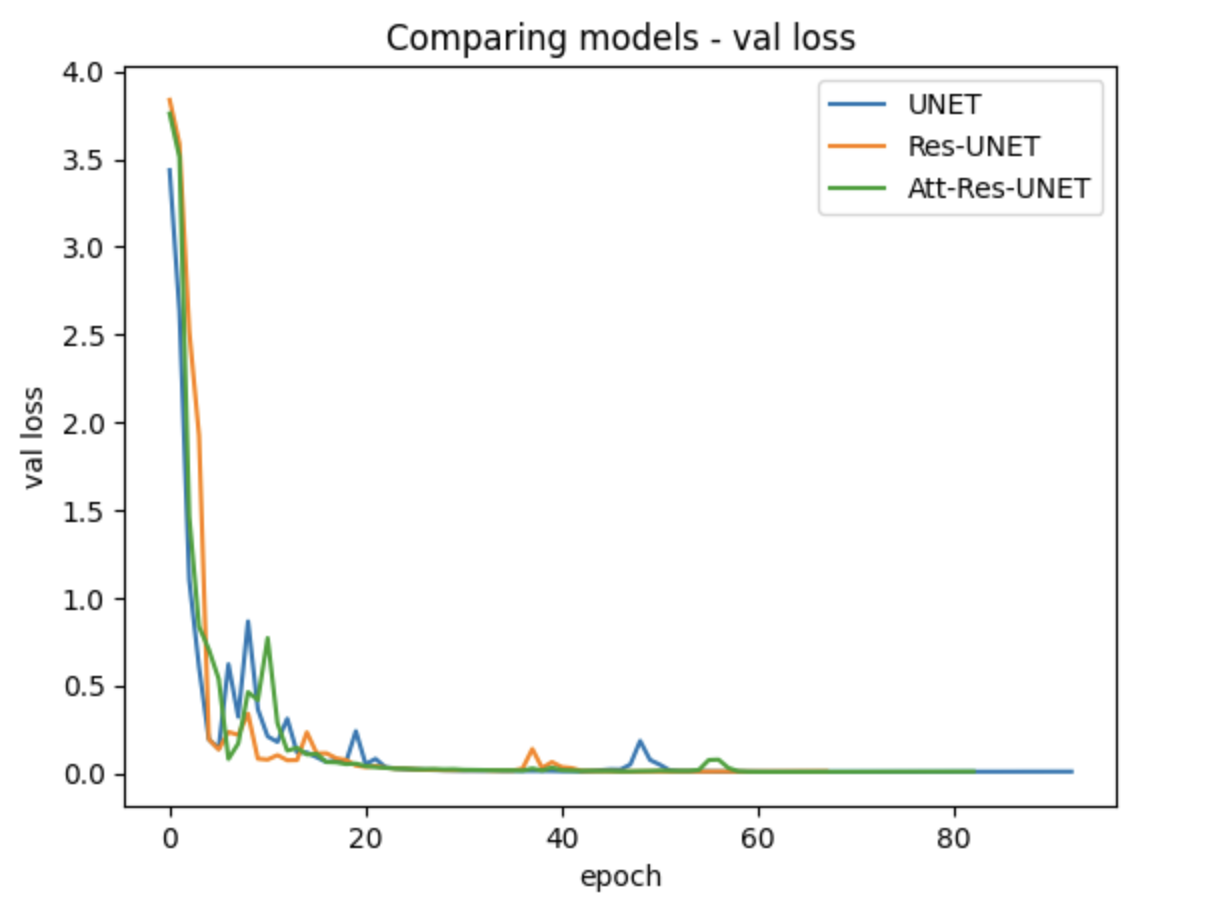}
    \caption{Convergence for trained models on Heart Segmentation}
    \label{fig: model_convergence_heart}
\end{figure}
Fig \ref{fig: model_convergence_heart} show the change in Categorical Focal Crossentropy over epochs for validation data for the three models. All models converge similarly, starting with a high initial loss that rapidly decreases within the first 10 epochs. Afterward, they exhibit noticeable fluctuations in loss, with Res-UNet showing fewer fluctuations compared to the others. Overall, their convergence patterns are similar.
\begin{table}[ht]
    \centering
    \caption{Precision and Recall score for each class by three models}
    \begin{tabular}{|c|c|c|c|c|c|c|c|c|}
        \hline
        & \multicolumn{4}{|c|}{Precision} & \multicolumn{4}{|c|}{Recall} \\
        \hline
        Models & 0 & 1 & 2 & 3 & 0 & 1 & 2 & 3 \\
        \hline
        UNet & 0.99 & 0.91 & 0.89 & \textbf{0.96} & 0.99 & 0.91 & \textbf{0.89} & 0.94 \\
        \hline
        Res-UNet & 0.99 & \textbf{0.92} & 0.89 & 0.95 & 0.99 & \textbf{0.92} & \textbf{0.89} & 0.94 \\
        \hline
        Attention Res-UNet & 0.99 & 0.91 & 0.89 & 0.94 & 0.99 & 0.91 & 0.88 & \textbf{0.95} \\
        \hline
    \end{tabular}
    \label{tab:heart_prec_recall}
\end{table}
\begin{table}[ht]
    \centering
    \caption{Dice and IoU score for each class by three models}
    \begin{tabular}{|c|c|c|c|c|c|c|c|c|}
        \hline
        & \multicolumn{4}{|c|}{Dice} & \multicolumn{4}{|c|}{IoU} \\
        \hline
        Models & 0 & 1 & 2 & 3 & 0 & 1 & 2 & 3 \\
        \hline
        UNet & 0.993 & 0.906 & \textbf{0.893} & \textbf{0.951} & \textbf{0.987} & 0.829 & \textbf{0.807} & \textbf{0.907} \\
        \hline
        Res-UNet & 0.993 & \textbf{0.92} & 0.888 & 0.944 & \textbf{0.987} & \textbf{0.852} & 0.799 & 0.895 \\
        \hline
        Attention Res-UNet & 0.993 & 0.908 & 0.884 & 0.945 & 0.985 & 0.831 & 0.792 & 0.895 \\
        \hline
    \end{tabular}
    \label{tab:heart_dice_iou}
\end{table}
\begin{table}[ht]
    \centering
    \caption{Accuracy and Loss score by the three models}
    \begin{tabular}{|c|c|c|}
        \hline
        Models & Accuracy & Loss \\
        \hline
        UNet & \textbf{98.41\%} & \textbf{1.00\%} \\
        \hline
        Res-UNet & \textbf{98.41\%} & 1.09\% \\
        \hline
        Attention Res-UNet & 98.28\% & 1.44\% \\
        \hline
    \end{tabular}
    \label{tab:heart_acc_loss}
\end{table}
Table \ref{tab:heart_prec_recall} provides precision and recall values for each class predicted by the three models.

\noindent
\textbf{Class-wise performance evaluation}
\begin{itemize}
    \item \textbf{Class 0 (Background):} All models achieve a high precision score, indicating that they are good at minimizing false positives for the background class. However, UNet and Res-UNet achieve the highest recall, suggesting that they capture most of the background pixels. UNet achieves the highest Dice coefficient and IoU, indicating its accuracy in identifying the background class.
    
    \item \textbf{Class 1 (RV Cavity):} Res-UNet achieves the highest precision for this class, indicating its accuracy in positive predictions. It also has the highest recall, meaning it captures most of the RV cavity pixels. Res-UNet has the highest Dice coefficient, indicating accurate spatial predictions, while UNet has the highest IoU.
    
    \item \textbf{Class 2 (Myocardium):} UNet has the highest precision for the myocardium, indicating accurate positive predictions, while Res-UNet has the highest recall, capturing the most myocardium pixels. UNet achieves the highest Dice coefficient and IoU for the myocardium.
    
    \item \textbf{Class 3 (LV Cavity):} Attention Res-UNet achieves the highest precision for the LV cavity, indicating its proficiency in minimizing false positives. It also has the highest recall, suggesting it captures most of the LV cavity pixels. UNet achieves the highest Dice coefficient and IoU for the LV cavity.
\end{itemize}

    
    
    

    
    
    

\noindent
\textbf{Accuracy and Loss}
\begin{itemize}
    \item UNet achieves the highest accuracy of 98.41\%, indicating its proficiency in overall pixel-level classification. UNet also has the lowest loss at 1.00\%, suggesting that it minimizes the difference between predicted and ground truth masks effectively.
    
    \item Res-UNet achieves a similar accuracy of 98.41\% but has a slightly higher loss at 1.09\%.
    
    \item Attention Res-UNet has an accuracy of 98.28\% and the highest loss at 1.44
\end{itemize}
In summary, the results show that each model excels in different aspects:
\begin{itemize}
    \item UNet demonstrates high accuracy, low loss, and strong performance in capturing background, myocardium, and LV cavity classes.
    \item Res-UNet achieves high precision and recall for the RV cavity and the highest Dice coefficient for class 1 (RV Cavity).
    \item Attention Res-UNet excels in precision and recall for the LV cavity and class 3 (LV Cavity).
\end{itemize}
\begin{figure}[h!]
  \centering
  \subfloat{\includegraphics[width=0.8\textwidth]{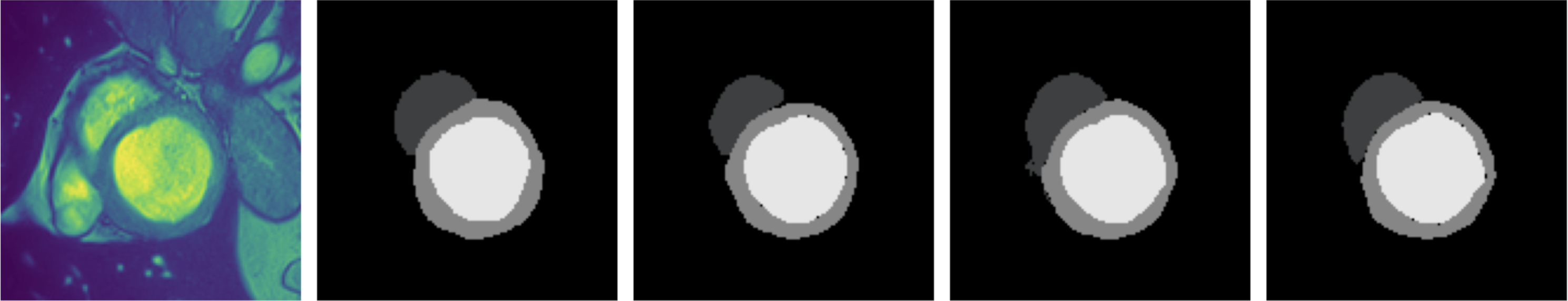}}\\
  \subfloat{\includegraphics[width=0.8\textwidth]{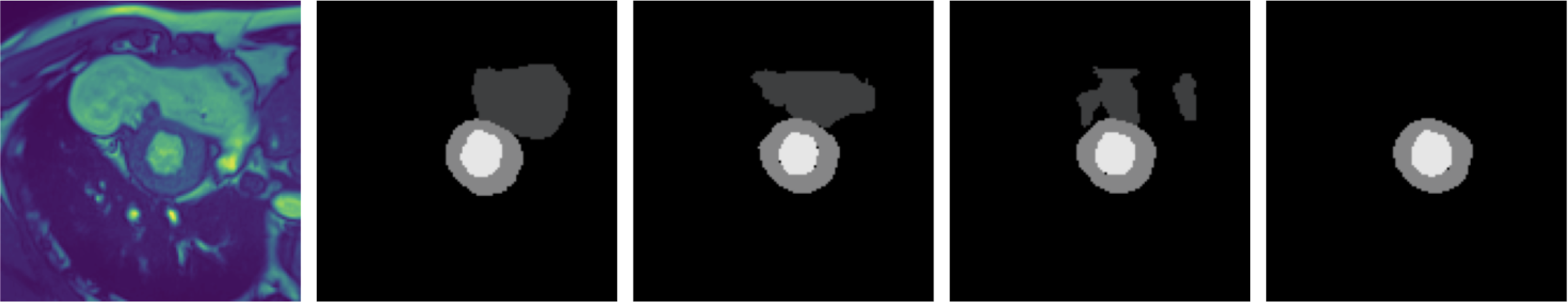}}\\
  \subfloat{\includegraphics[width=0.8\textwidth]{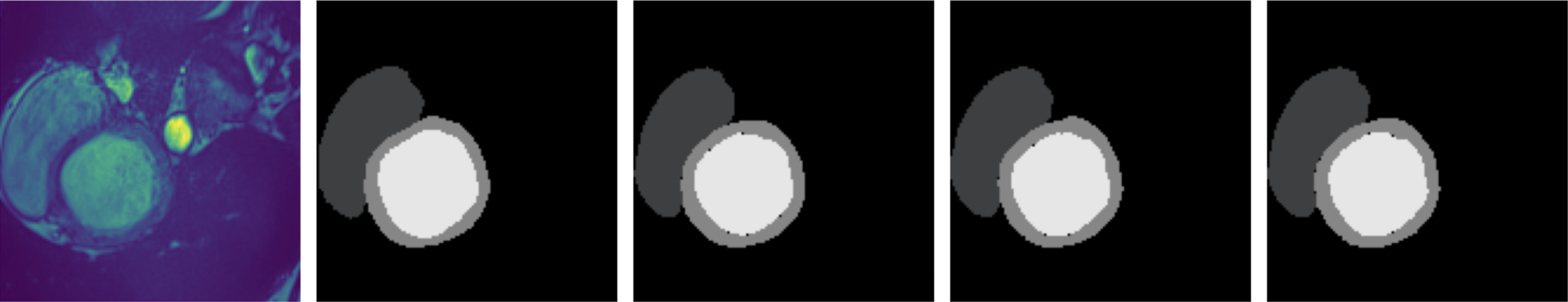}}\\
  \subfloat{\includegraphics[width=0.8\textwidth]{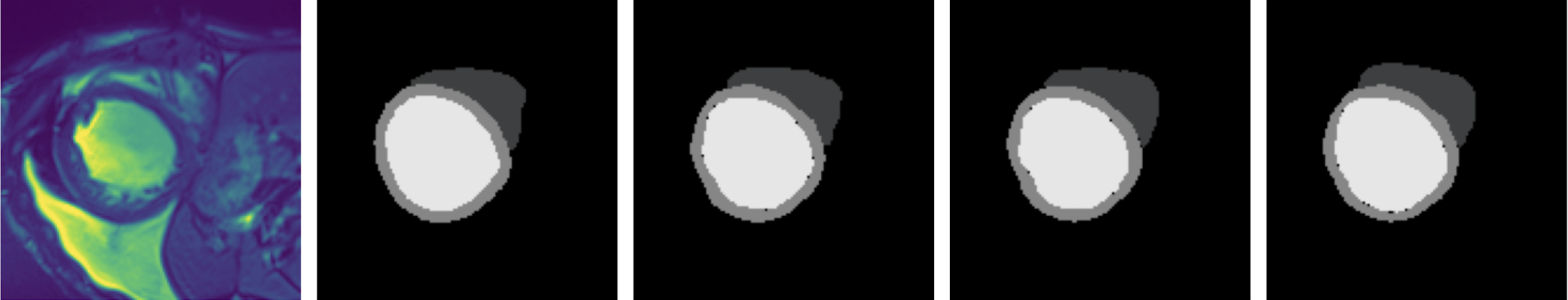}}
  \caption{Segmentation results by the three models for four different examples, from left to right are the input images, ground-truth, segmentation results by UNet, Res-UNet and Attention Res-UNet.}
  \label{fig: heart_models_res}
\end{figure}
\subsubsection{Discussion}
Figure \ref{fig: heart_models_res} shows four examples with given image and ground truth mask followed my predictions from the three models.

The trained models (UNet, Res-UNet, and Attention Res-UNet) exhibit acceptable results in producing masks similar to the ground truth for the multi-class image segmentation task involving the classes Myocardium (Class 2), LV cavity (Class 3), and RV cavity. 

All three models generally perform well in producing accurate image segmentation masks, particularly for the Myocardium (Class 2) and LV cavity (Class 3) due to their abundance of training examples and distinctive features. However, they struggle with the underrepresented RV cavity class (Class 1), resulting in frequent misclassifications, likely due to the limited training data for this class.

Despite these challenges, UNet outperforms the other models in capturing Class 1 pixels. This may be attributed to UNet's lower focal loss score for Class 0, indicating its better handling of class imbalance and enhanced focus on the RV cavity class.

In summary, the models encounter typical challenges associated with imbalanced class distributions in multi-class image segmentation. They excel with well-represented classes but face difficulties with underrepresented ones. UNet shows promise in handling this imbalance, but further improvements, such as addressing class balance and using data augmentation, could enhance performance across all classes.

\section{Conclusions}
We have evaluated the performance of UNet, Res-UNet and Attention Res-UNet on three problems of  brain tumor, polyp and multi-label heart segmentation. All models achieved acceptable segmentation results when compared to the ground-truth provided by the datasets.
Differences were visible when the target masks become more complex in nature. The key findings of the study are summarised as follows:
\begin{enumerate}
    \item UNet often misclassified target classes as background when overall target pixel is relatively small such as brain tumor  or small polyp segments. UNet also struggled with target segmentation when mask edge boundaries were intricate in nature. This pointed to UNet's limitations with vanishing gradient problem and inability to put focus on hard-to-classify pixels. 
    \item Res-Unet and Attention Res-Unet proved to be more suitable in handling complex and irregular structures, as both the models were able to capture the complex boundaries in most cases. This was indicative of the residual connection introduced in the two models, which mitigated the vanishing gradient problem. 
    \item Attention Res-Unet was more effective at tackling class imbalance, as it consistently achieved high recall values among all all tasks. The model was also predicted more refined masks in most cases compared to Res-UNet. Multi-label heart segmentation enforced these theories, as the mask images were less imbalanced compared to other tasks and resulted in higher performance in standard UNet model. Res-Unet and Attention Res-UNet performed similarly in this task due to exclusion of major class under-representation in most. One of the three classes was often misclassified due to its scarcity in most of the images in the dataset. This indicated that datasets need to be more inclusive of all classes in order for these robust models to perform at their full potential. 
\end{enumerate}

The implications of this work extend beyond the immediate research domain. It sets a modern benchmark for segmentation techniques in the medical field, offering future researchers valuable insights into the critical factors to consider when applying UNet, variants and other deep learning methodologies to medical image analysis. To enhance this study, future work could focus on the application of the aforementioned models to three-dimensional medical images, as many medical datasets are inherently three-dimensional. Additionally, involving medical specialists to evaluate segmentation outputs could provide more refined and clinically relevant assessments. More loss functions and their effect on these models can be explored, thus adding to the reliability of the study and these models. Similar studies into more extensions of UNet and their suitability can also be explored, hence enriching concrete guidelines. 

\bibliographystyle{abbrv}
\bibliography{references}

\begin{thebibliography}{10}

\bibitem{aichinger2012radiation}
H.~Aichinger, J.~Dierker, S.~Joite-Barfu{\ss}, and M.~S{\"a}bel.
\newblock {\em Radiation exposure and image quality in X-ray diagnostic radiology: physical principles and clinical applications}.
\newblock Springer, 2012.

\bibitem{badrinarayanan2017segnet}
V.~Badrinarayanan, A.~Kendall, and R.~Cipolla.
\newblock Segnet: A deep convolutional encoder-decoder architecture for image segmentation.
\newblock {\em IEEE transactions on pattern analysis and machine intelligence}, 39(12):2481--2495, 2017.

\bibitem{Polyp_dataset}
J.~Bernal, F.~J. S{\'a}nchez, G.~Fern{\'a}ndez-Esparrach, D.~Gil, C.~Rodr{\'\i}guez, and F.~Vilari{\~n}o.
\newblock Wm-dova maps for accurate polyp highlighting in colonoscopy: Validation vs. saliency maps from physicians.
\newblock {\em Computerized medical imaging and graphics : the official journal of the Computerized Medical Imaging Society}, 43:99---111, July 2015.

\bibitem{ACDC_dataset}
O.~Bernard, A.~Lalande, C.~Zotti, F.~Cervenansky, X.~Yang, P.-A. Heng, I.~Cetin, K.~Lekadir, O.~Camara, M.~Ã. Gonzalez~Ballester, G.~Sanroma, S.~Napel, S.~Petersen, G.~Tziritas, G.~Ilias, M.~Khened, V.~Kollerathu, G.~Krishnamurthi, M.-M. Rohe, and P.-M. Jodoin.
\newblock Deep learning techniques for automatic mri cardiac multi-structures segmentation and diagnosis: Is the problem solved?
\newblock {\em IEEE Transactions on Medical Imaging}, PP:1--1, 05 2018.

\bibitem{beucher1992watershed}
S.~Beucher.
\newblock The watershed transformation applied to image segmentation.
\newblock {\em Scanning microscopy}, 1992(6):28, 1992.

\bibitem{brain_data}
M.~Buda.
\newblock Brain {MRI} segmentation, 2017.
\newblock https://www.kaggle.com/datasets/mateuszbuda/lgg-mri-segmentation.

\bibitem{callaghan2015evaluation}
M.~F. Callaghan, O.~Josephs, M.~Herbst, M.~Zaitsev, N.~Todd, and N.~Weiskopf.
\newblock An evaluation of prospective motion correction (pmc) for high resolution quantitative mri.
\newblock {\em Frontiers in neuroscience}, 9:97, 2015.

\bibitem{chen2018novel}
J.~Chen, H.~Zheng, X.~Lin, Y.~Wu, and M.~Su.
\newblock A novel image segmentation method based on fast density clustering algorithm.
\newblock {\em Engineering Applications of Artificial Intelligence}, 73:92--110, 2018.

\bibitem{cootes1995active}
T.~F. Cootes, C.~J. Taylor, D.~H. Cooper, and J.~Graham.
\newblock Active shape models-their training and application.
\newblock {\em Computer vision and image understanding}, 61(1):38--59, 1995.

\bibitem{Dodo:bioimaging2018}
B.~Dodo, Y.~Li, K.~Eltayef, and X.~Liu.
\newblock Graph-cut segmentation of retinal layers from oct images.
\newblock In {\em International Conference on Bioimaging}, 2018.

\bibitem{Dodo:bioimaging2018_2}
B.~Dodo, Y.~Li, K.~Eltayef, and X.~Liu.
\newblock Min-cut segmentation of retinal oct images.
\newblock In {\em International Joint Conference on Biomedical Engineering Systems and Technologies}, pages 86--99. Springer, 2018.

\bibitem{Dodo:jms2019}
B.~Dodo, Y.~Li, K.~Eltayef, and X.~Liu.
\newblock Automatic annotation of retinal layers in optical coherence tomography images.
\newblock {\em Journal of Medical Systems.}, 2019.

\bibitem{Dodo:best2019}
B.~Dodo, Y.~Li, K.~Eltayef, and X.~Liu.
\newblock Min-cut segmentation of retinal oct images.
\newblock In {\em Cliquet Jr. A. et al. (eds) Biomedical Engineering Systems and Technologies}. Springer., 2019.

\bibitem{Dodo:cbms2017}
B.~Dodo, Y.~Li, and X.~Liu.
\newblock Retinal oct image segmentation using fuzzy histogram hyperbolization and continuous max-flow.
\newblock In {\em IEEE International Symposium on Computer-Based Medical Systems}, 2017.

\bibitem{Dodo:bioimaging2019}
B.~Dodo, Y.~Li, X.~Liu, and M.~Dodo.
\newblock Level set segmentation of retinal oct images.
\newblock In {\em International Conference on Bioimaging. Czech Republic.}, 2019.

\bibitem{Dodo:cbms2019}
B.~Dodo, Y.~Li, A.~Tucker, D.~Kaba, and X.~Liu.
\newblock Retinal oct segmentation using fuzzy region competition and level set methods.
\newblock In {\em 2019 IEEE 32nd International Symposium on Computer-Based Medical Systems (CBMS)}, pages 93--98. IEEE, 2019.

\bibitem{Dodo:access2019}
B.~I. Dodo, Y.~Li, D.~Kaba, and X.~Liu.
\newblock Retinal layer segmentation in optical coherence tomography images.
\newblock {\em IEEE Access}, 7:152388--152398, 2019.

\bibitem{drozdzal2018learning}
M.~Drozdzal, G.~Chartrand, E.~Vorontsov, M.~Shakeri, L.~Di~Jorio, A.~Tang, A.~Romero, Y.~Bengio, C.~Pal, and S.~Kadoury.
\newblock Learning normalized inputs for iterative estimation in medical image segmentation.
\newblock {\em Medical image analysis}, 44:1--13, 2018.

\bibitem{duarte2020speckle}
C.~A. Duarte-Salazar, A.~E. Castro-Ospina, M.~A. Becerra, and E.~Delgado-Trejos.
\newblock Speckle noise reduction in ultrasound images for improving the metrological evaluation of biomedical applications: an overview.
\newblock {\em IEEE Access}, 8:15983--15999, 2020.

\bibitem{Eltayef:ida2017}
K.~Eltayef, Y.~Li, B.~Dodo, and X.~Liu.
\newblock Skin cancer detection in dermoscopy images using sub-region features.
\newblock In {\em International Symposium on Intelligent Data Analysis}, pages 75--86. Springer, Cham, 2017.

\bibitem{eltayef2016detection1}
K.~Eltayef, Y.~Li, and X.~Liu.
\newblock Detection of melanoma skin cancer in dermoscopy images.
\newblock In {\em In Proc. International Conference on Communication, Image and Signal Processing. Dubai, UAE, .}, 2016.

\bibitem{eltayef2016detection2}
K.~Eltayef, Y.~Li, and X.~Liu.
\newblock Detection of pigment networks in dermoscopy images.
\newblock In {\em In Proc. International Conference on Communication, Image and Signal Processing. Dubai, UAE, .}, 2016.

\bibitem{Eltayef:cbms2017}
K.~Eltayef, Y.~Li, and X.~Liu.
\newblock Lesion segmentation in dermoscopy images using particle swarm optimization and markov random field.
\newblock In {\em IEEE International Symposium on Computer-Based Medical Systems}, 2017.

\bibitem{felzenszwalb2004efficient}
P.~F. Felzenszwalb and D.~P. Huttenlocher.
\newblock Efficient graph-based image segmentation.
\newblock {\em International journal of computer vision}, 59:167--181, 2004.

\bibitem{fraz2012blood}
M.~M. Fraz, P.~Remagnino, A.~Hoppe, B.~Uyyanonvara, A.~R. Rudnicka, C.~G. Owen, and S.~A. Barman.
\newblock Blood vessel segmentation methodologies in retinal images--a survey.
\newblock {\em Computer methods and programs in biomedicine}, 108(1):407--433, 2012.

\bibitem{fritscher2014automatic}
K.~D. Fritscher, M.~Peroni, P.~Zaffino, M.~F. Spadea, R.~Schubert, and G.~Sharp.
\newblock Automatic segmentation of head and neck ct images for radiotherapy treatment planning using multiple atlases, statistical appearance models, and geodesic active contours.
\newblock {\em Medical physics}, 41(5):051910, 2014.

\bibitem{haque2020deep}
I.~R.~I. Haque and J.~Neubert.
\newblock Deep learning approaches to biomedical image segmentation.
\newblock {\em Informatics in Medicine Unlocked}, 18:100297, 2020.

\bibitem{havaei2017brain}
M.~Havaei, A.~Davy, D.~Warde-Farley, A.~Biard, A.~Courville, Y.~Bengio, C.~Pal, P.-M. Jodoin, and H.~Larochelle.
\newblock Brain tumor segmentation with deep neural networks.
\newblock {\em Medical image analysis}, 35:18--31, 2017.

\bibitem{he2015deep}
K.~He, X.~Zhang, S.~Ren, and J.~Sun.
\newblock Deep residual learning for image recognition, 2015.

\bibitem{huda2003review}
W.~Huda and R.~M. Slone.
\newblock {\em Review of radiologic physics}.
\newblock Lippincott Williams \& Wilkins, 2003.

\bibitem{iglesias2011robust}
J.~E. Iglesias, C.-Y. Liu, P.~M. Thompson, and Z.~Tu.
\newblock Robust brain extraction across datasets and comparison with publicly available methods.
\newblock {\em IEEE transactions on medical imaging}, 30(9):1617--1634, 2011.

\bibitem{isensee2021nnu}
F.~Isensee, P.~F. Jaeger, S.~A. Kohl, J.~Petersen, and K.~H. Maier-Hein.
\newblock nnu-net: a self-configuring method for deep learning-based biomedical image segmentation.
\newblock {\em Nature methods}, 18(2):203--211, 2021.

\bibitem{jain1988algorithms}
A.~K. Jain and R.~C. Dubes.
\newblock {\em Algorithms for clustering data}.
\newblock Prentice-Hall, Inc., 1988.

\bibitem{Kaba:his2013}
D.~Kaba, A.~G. Salazar-Gonzalez, Y.~Li, X.~Liu, and A.~Serag.
\newblock Segmentation of retinal blood vessels using gaussian mixture models and expectation maximisation.
\newblock In {\em International Conference on Health Information Science}, pages 105--112, 2013.

\bibitem{Kaba:hiss2014}
D.~Kaba, C.~Wang, Y.~Li, A.~Salazar-Gonzalez, X.~Liu, and A.~Serag.
\newblock Retinal blood vessels extraction using probabilistic modelling.
\newblock {\em Health Information Science and Systems}, 2(1):2, 2014.

\bibitem{Kaba:oe2015}
D.~Kaba, Y.~Wang, C.~Wang, X.~Liu, H.~Zhu, A.~G. Salazar-Gonzalez, and Y.~Li.
\newblock Retina layer segmentation using kernel graph cuts and continuous max-flow.
\newblock {\em Optics Express}, 23(6):7366--7384, 2015.

\bibitem{kamiyoshihara2004congenital}
M.~Kamiyoshihara, A.~Otaki, T.~Nameki, O.~Kawashima, Y.~Otani, K.~Sakata, and Y.~Morishita.
\newblock Congenital bronchial atresia treated with video-assisted thoracoscopic surgery; report of a case.
\newblock {\em Kyobu geka. The Japanese Journal of Thoracic Surgery}, 57(7):591--593, 2004.

\bibitem{kao2019brain}
P.-Y. Kao, T.~Ngo, A.~Zhang, J.~W. Chen, and B.~Manjunath.
\newblock Brain tumor segmentation and tractographic feature extraction from structural mr images for overall survival prediction.
\newblock In {\em Brainlesion: Glioma, Multiple Sclerosis, Stroke and Traumatic Brain Injuries: 4th International Workshop, BrainLes 2018, Held in Conjunction with MICCAI 2018, Granada, Spain, September 16, 2018, Revised Selected Papers, Part II 4}, pages 128--141. Springer, 2019.

\bibitem{kass1988snakes}
M.~Kass, A.~Witkin, and D.~Terzopoulos.
\newblock Snakes: Active contour models.
\newblock {\em International journal of computer vision}, 1(4):321--331, 1988.

\bibitem{litjens2017survey}
G.~Litjens, T.~Kooi, B.~E. Bejnordi, A.~A.~A. Setio, F.~Ciompi, M.~Ghafoorian, J.~A. Van Der~Laak, B.~Van~Ginneken, and C.~I. S{\'a}nchez.
\newblock A survey on deep learning in medical image analysis.
\newblock {\em Medical image analysis}, 42:60--88, 2017.

\bibitem{liu2020survey}
L.~Liu, J.~Cheng, Q.~Quan, F.-X. Wu, Y.-P. Wang, and J.~Wang.
\newblock A survey on u-shaped networks in medical image segmentations.
\newblock {\em Neurocomputing}, 409:244--258, 2020.

\bibitem{lock2022factors}
F.~K. Lock and D.~Carrieri.
\newblock Factors affecting the uk junior doctor workforce retention crisis: an integrative review.
\newblock {\em BMJ open}, 12(3):e059397, 2022.

\bibitem{ma2017cascade}
J.~Ma, F.~Wu, T.~Jiang, J.~Zhu, and D.~Kong.
\newblock Cascade convolutional neural networks for automatic detection of thyroid nodules in ultrasound images.
\newblock {\em Medical physics}, 44(5):1678--1691, 2017.

\bibitem{MAJI_attention}
D.~Maji, P.~Sigedar, and M.~Singh.
\newblock Attention res-unet with guided decoder for semantic segmentation of brain tumors.
\newblock {\em Biomedical Signal Processing and Control}, 71:103077, 2022.

\bibitem{mcconnell2022integrating}
N.~McConnell, A.~Miron, Z.~Wang, and Y.~Li.
\newblock Integrating residual, dense, and inception blocks into the nnunet.
\newblock In {\em IEEE 35th International Symposium on Computer Based Medical Systems}, 2022.

\bibitem{mcconnel2023}
N.~McConnell, N.~Ndipenoch, Y.~Cao, A.~Miron, and Y.~Li.
\newblock Advanced architectural variations of nnunet.
\newblock {\em Neurocomputing}, 2023.

\bibitem{ndipenoch2022simultaneous}
N.~Ndipenoch, A.~Miron, Z.~Wang, and Y.~Li.
\newblock Simultaneous segmentation of layers and fluids in retinal oct images.
\newblock In {\em IEEE Conference on Image and Signal Processing, BioMedical Engineering and Informatics}, 2022.

\bibitem{ndipenoch2023retinal}
N.~Ndipenoch, A.~Miron, Z.~Wang, and Y.~Li.
\newblock Retinal image segmentation with small datasets.
\newblock In {\em 10th International Conference on Bioimaging}, 2023.

\bibitem{pedano2016tcga}
N.~Pedano, A.~E. Flanders, L.~Scarpace, T.~Mikkelsen, J.~M. Eschbacher, B.~Hermes, V.~Sisneros, J.~Barnholtz-Sloan, and Q.~Ostrom.
\newblock The cancer genome atlas low grade glioma collection (tcga-lgg) (version 3).
\newblock The Cancer Imaging Archive, 2016.

\bibitem{plenge2012super}
E.~Plenge, D.~H. Poot, M.~Bernsen, G.~Kotek, G.~Houston, P.~Wielopolski, L.~van~der Weerd, W.~J. Niessen, and E.~Meijering.
\newblock Super-resolution methods in mri: can they improve the trade-off between resolution, signal-to-noise ratio, and acquisition time?
\newblock {\em Magnetic resonance in medicine}, 68(6):1983--1993, 2012.

\bibitem{rabbani2008speckle}
H.~Rabbani, M.~Vafadust, P.~Abolmaesumi, and S.~Gazor.
\newblock Speckle noise reduction of medical ultrasound images in complex wavelet domain using mixture priors.
\newblock {\em IEEE transactions on biomedical engineering}, 55(9):2152--2160, 2008.

\bibitem{ren2015faster}
S.~Ren, K.~He, R.~Girshick, and J.~Sun.
\newblock Faster r-cnn: Towards real-time object detection with region proposal networks.
\newblock {\em Advances in neural information processing systems}, 28, 2015.

\bibitem{ronneberger2015unet}
O.~Ronneberger, P.~Fischer, and T.~Brox.
\newblock U-net: Convolutional networks for biomedical image segmentation, 2015.

\bibitem{Salazar:jbhi2014}
A.~Salazar-Gonzalez, D.~Kaba, Y.~Li, and X.~Liu.
\newblock Segmentation of the blood vessels and optic disk in retinal images.
\newblock {\em IEEE journal of biomedical and health informatics}, 18(6):1874--1886, 2014.

\bibitem{Salazar:his2012}
A.~Salazar-Gonzalez, Y.~Li, and D.~Kaba.
\newblock \uppercase{MRF} reconstruction of retinal images for the optic disc segmentation.
\newblock In {\em International Conference on Health Information Science}, pages 88--99, 2012.

\bibitem{Salazar:jaiscr2012}
A.~Salazar-Gonzalez, Y.~Li, and X.~Liu.
\newblock Automatic graph cut based segmentation of retinal optic disc by incorporating blood vessel compensation.
\newblock {\em Journal of Artificial Intelligence and Soft Computing Research}, 2(3):235--245, 2012.

\bibitem{Salazar:icarcv2010}
A.~G. Salazar-Gonzalez, Y.~Li, and X.~Liu.
\newblock Retinal blood vessel segmentation via graph cut.
\newblock In {\em International Conference on Control Automation Robotics \& Vision}, pages 225--230, 2010.

\bibitem{Salazar:cimi2011}
A.~G. Salazar-Gonzalez, Y.~Li, and X.~Liu.
\newblock Optic disc segmentation by incorporating blood vessel compensation.
\newblock In {\em IEEE Third International Workshop On Computational Intelligence In Medical Imaging}, pages 1--8, 2011.

\bibitem{sezgin2004survey}
M.~Sezgin and B.~l. Sankur.
\newblock Survey over image thresholding techniques and quantitative performance evaluation.
\newblock {\em Journal of Electronic imaging}, 13(1):146--168, 2004.

\bibitem{valvano2019convolutional}
G.~Valvano, G.~Santini, N.~Martini, A.~Ripoli, C.~Iacconi, D.~Chiappino, D.~Della~Latta, et~al.
\newblock Convolutional neural networks for the segmentation of microcalcification in mammography imaging.
\newblock {\em Journal of healthcare engineering}, 2019, 2019.

\bibitem{Wang:jms2015}
C.~Wang, D.~Kaba, and Y.~Li.
\newblock Level set segmentation of optic discs from retinal images.
\newblock {\em Journal of Medical Systems}, 4(3):213--220, 2015.

\bibitem{wang2020blood}
C.~Wang and Y.~Li.
\newblock Blood vessel segmentation from retinal images.
\newblock In {\em The 20th IEEE International Conference on BioInformatics And BioEngineering}, 2020.

\bibitem{Wang:icig2015}
C.~Wang, Y.~Wang, D.~Kaba, Z.~Wang, X.~Liu, and Y.~Li.
\newblock Automated layer segmentation of 3d macular images using hybrid methods.
\newblock In {\em Proc. International Conference on Image and Graphics. Tianjing, China.}, volume 9217, pages 614--628, 2015.

\bibitem{Wang:icig2015_2}
C.~Wang, Y.~Wang, D.~Kaba, H.~Zhu, Z.~Wang, X.~Liu, and Y.~Li.
\newblock Segmentation of intra-retinal layers in 3d optic nerve head images.
\newblock In {\em Proc. International Conference on Image and Graphics. Tianjing.}, volume 9219, pages 321--332, 2015.

\bibitem{Wang:jbhi2017}
C.~Wang, Y.~Wang, and Y.~Li.
\newblock Automatic choroidal layer segmentation using markov random field and level set method.
\newblock {\em IEEE journal of biomedical and health informatics}, 2017.

\bibitem{wells2011medical}
P.~N. Wells and H.-D. Liang.
\newblock Medical ultrasound: imaging of soft tissue strain and elasticity.
\newblock {\em Journal of the Royal Society Interface}, 8(64):1521--1549, 2011.

\bibitem{zaitsev2015motion}
M.~Zaitsev, J.~Maclaren, and M.~Herbst.
\newblock Motion artifacts in mri: A complex problem with many partial solutions.
\newblock {\em Journal of Magnetic Resonance Imaging}, 42(4):887--901, 2015.

\end{thebibliography}

\end{document}